# Hyperspectral near infrared imaging using a tunable spectral phasor

Jan Stegemann[1,2], Franziska Gröniger[2], Krisztian Neutsch[1], Han Li[3,4], Benjamin Flavel[5], Justus Tom Metternich[1,2], Luise Erpenbeck[6], Poul Petersen[1], Per Niklas Hedde[7], Sebastian Kruss[1,2*]

[1]Department of Chemistry and Biochemistry, Bochum University, Bochum, Germany

[2]Fraunhofer Institute for Microelectronic Circuits and Systems, Duisburg, Germany

[3]Department of Mechanical and Materials Engineering, University of Turku, FI-20014 Turku, Finland

[4]Turku Collegium for Science, Medicine and Technology, University of Turku, FI-20520 Turku, Finland

[5]Institute of Nanotechnology, Karlsruhe Institute of Technology, Karlsruhe, Germany

[6]Department of Dermatology, University Hospital Münster, Münster, Germany

[7]Beckman Laser Institute & Medical Clinic, University of California, Irvine, CA, USA

Corresponding author: sebastian.kruss@rub.de

## Abstract

Hyperspectral imaging captures both spectral and spatial information from a sample. The near infrared (NIR, > 800 nm) is advantageous for biomedical imaging as it falls into the tissue transparency window but also contains vibrational overtone and combination modes useful for molecular fingerprinting. Here, we demonstrate hyperspectral NIR imaging using a spectral phasor transformation (HyperNIR). This method employs a liquid crystal variable retarder (LCVR) for tunable, wavelength-dependent sine-, cosine and no filtering that transforms optical signals into phasor space. Spectral information is thus obtained with just three images. The LCVR can be adjusted to cover a spectral range from 900 nm to 1600 nm in windows tunable from 50 nm to 700 nm. This approach enables distinguishing NIR fluorophores with emission peaks less than 5 nm apart. Furthermore, we demonstrate label-free hyperspectral NIR reflectance imaging to identify plastic polymers and to monitor *in vivo* plant health. The approach uses the full camera resolution and reaches hyperspectral frame rates of 0.2 per second, limited only by the switching rate of the LCVR. HyperNIR facilitates straightforward hyperspectral imaging with standard NIR cameras for applications in biomedical imaging and environmental monitoring.

# Introduction

Label-free imaging techniques enable non-destructive and high-resolution interrogation of materials and biomedical samples without the need for exogenous labels[1]. These techniques leverage inherent optical properties of samples to generate contrast and extract detailed information about their structure, composition, and function. Prominent label-free imaging modalities include phase contrast[2,3], polarization[4], infrared-absorption[5], autofluorescence[6], photoacoustic[7,8], scattering-based[9,10,11] or Raman[1] microscopy. Such techniques are employed not only for basic research but also clinical diagnostics[1] and drug development[12]. In materials science, these techniques contribute to the characterization of materials at the micro- and nano-scale, aiding in the design and optimization of novel materials for diverse applications[13]. Despite many advances weak signals, limited penetration depth and specialized instrumentation limit the use of these techniques[1].

The near-infrared (NIR) range of the electromagnetic spectrum (800 nm – 2500 nm) corresponds mainly to overtone and combinational vibrational modes. NIR spectroscopy exploits these modes to characterize samples rapidly and non-destructively[14]. Measuring the absorption, reflection, or transmission of NIR light provides quantitative or qualitative information about chemical composition, structure, and physical properties of materials. This analytical technique finds widespread use for pharmaceuticals[15], in agriculture[16,17] and food[18] industries, and in biomedical research[19,20]. The NIR is also highly beneficial for biomedical applications due to the so called "optical windows" from 650 to 1350 nm and above 1450 nm[21]. As absorption and scattering are reduced in this region, light can penetrate deeper into biological tissue than visible (Vis) and mid-IR light. Ongoing developments in instrumentation and data analysis methods (chemometrics) enhance the capabilities and applicability of NIR spectroscopy[14]. This progress is complemented by the development of novel NIR fluorescent nanomaterials[22]. They enable NIR fluorescence imaging with high signal to noise ratios. One example are single-walled carbon nanotubes (SWCNT) that can be chemically tailored as biosensors for detection and imaging of signalling molecules, pathogens or disease markers[23,24,25,26].

Combining spectral and spatial information (hyperspectral imaging) enables novel applications such as environmental monitoring using drones[27]. In contrast to conventional imaging, hyperspectral imaging assigns a spectrum or at least spectral information to each pixel creating a 3D data cube (Fig. 1a). Hyperspectral imaging is performed by either scanning spatially (pushbroom scanning) or spectrally[28], which makes it inherently slow. For the NIR serial imaging using volume Bragg gratings has been demonstrated[29]. An alternative are snapshot cameras that sacrifice spatial resolution. Cameras consisting of a Bayer-like mosaic pattern[30] or utilizing the light field technology[31,32] can achieve high framerates with a compromise between spectral and spatial resolution. The number of pixels in NIR (InGaAs) cameras is typically very low (around 0.1 Megapixel), thus keeping all camera pixels for a high resolution is desired. Other methods like Fourier coded aperture transform spectral imaging (FCTS) can compensate for spectral resolution, but the optical setup gets more sophisticated and extensive computational reconstruction is necessary[33].

One alternative is to determine a spectral phasor for each pixel, which maps each pixel into a 2D (phasor) space and the position contains the spectral information[34]. The phasor approach is well known for interpreting fluorescence lifetime images. It provides a fit-free analysis of time-resolved data, which simplifies the analysis of complex samples[35]. There is also a counterpart in the spectral domain, which requires images acquired at different wavelengths or the full spectrum of each pixel[34]. Instead of computing this transformation from a full spectrum, the same information can be obtained by spectral filtering with appropriate (hardware) filters[36]. Each pixel is assigned to a point in a phasor space, which indicates a specific wavelength or spectral signature. For this approach, commercially available absorption filters with a poor fit to sine/cosine[34] or custom-built interference filters were employed in the Vis range of the electromagnetic spectrum[37,38]. These filters are made for a fixed wavelength range and cannot be adapted to different spectral regions and ranges. Overall, the potential of using phasors has not been leveraged yet for hyperspectral imaging and the NIR range of the spectrum is especially challenging.

Here, we show a tunable and straightforward implementation of NIR hyperspectral imaging (Fig. 1b) of fluorescence, transmission or reflection signals from highly relevant samples.

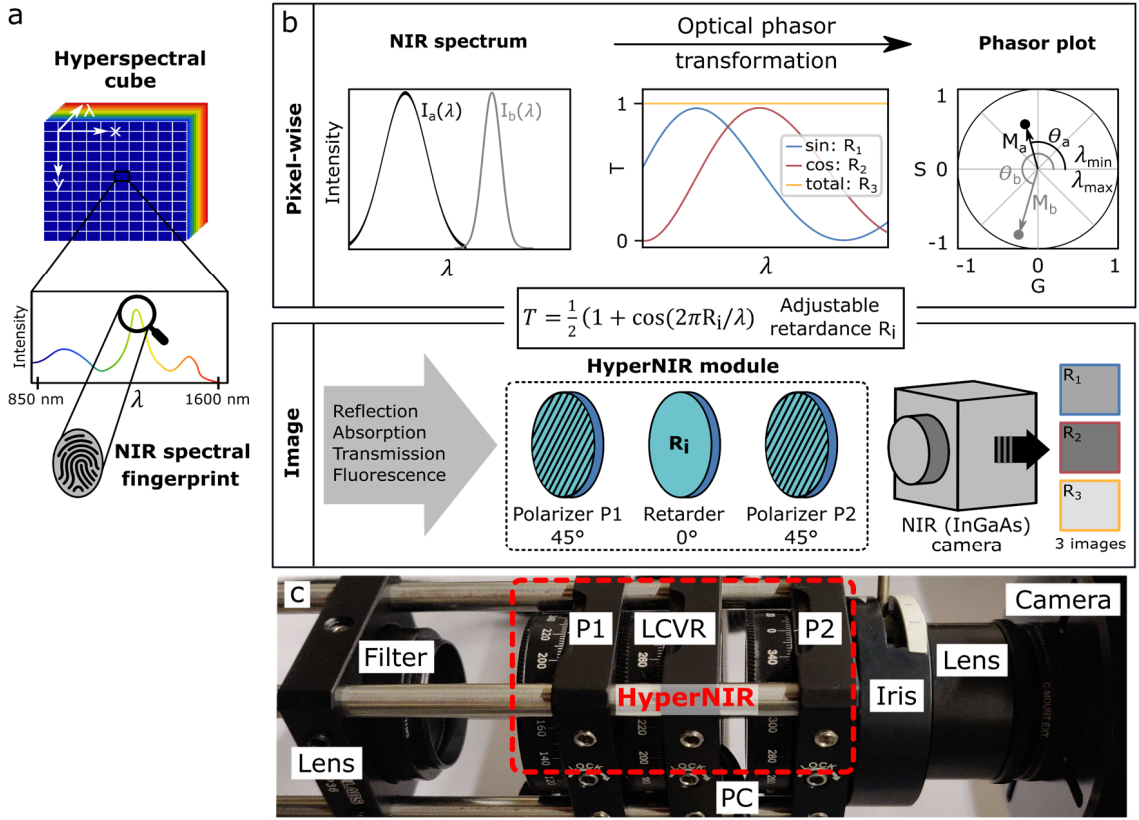

*Fig. 1: Implementation of phasor-based tunable hyperspectral NIR imaging.* a) A hyperspectral image corresponds to a 3D data cube (x,y,λ). In the near infrared (NIR) it contains either chemical information from combination and overtone modes that are useful for fingerprinting molecules/materials or signals from fluorophores. b) The spectral signal $I_i(\lambda)$ is transformed into the phasor space by measuring its intensity with a sine-, a cosine-shaped and total transmission filter $T(\lambda)$. Thereby, spectral information of a single pixel is represented as a point in 2D phasor space (G = cosine component of the spectrum, S = sine component of the spectrum, $M_i$ = modulation, $\theta_i$ = phase). For a full hyperspectral image, the NIR signals are transmitted through a tunable cosine/sine filter, which is generated by the HyperNIR module. It consists of a modular retarder between two parallel linear polarizers, which approximates cosine/sine/total transmission filters by changing the retardance. The covered spectral range $\lambda_{min} - \lambda_{max}$ is tunable because the retardance can be adjusted to squeeze or extend the sine/cosine. Three intensity images are taken by an InGaAs camera at different retardance ($R_1, R_2, R_3$) to obtain the phasor position for each single pixel of the image. c) Picture of HyperNIR module connected to a NIR (InGaAs) camera (P1 = polarizer 1, P2 = polarizer 2, LCVR = liquid crystal variable retarder, PC = connection to computer).

## Results

### Implementation of HyperNIR

To implement a tunable wavelength range, we used polarization optics to generate cosine- or sine-shaped spectral transmissions. The optical module (HyperNIR) consists of two parallel linear polarizers and a variable retarder (Fig. 1b). Both linear polarizers operate at an angle of 45° with respect to the fast axis of the variable retarder. NIR light from fluorescence, reflected or transmitted signals reaches the first polarizer. In most cases, this light is randomly polarized but could be partially polarized due to polarization dependent reflection, absorption or slow orientational dynamics of fluorophores. In either case, the first linear polarizer filters the 45° component. After the first polarizer, the retarder imposes a variable retardation, which creates an elliptically polarized beam depending on the wavelength. With the second linear polarizer, the 45° component of the elliptically polarized light is filtered out. The overall effect of the two polarizers sandwiched around the retarder is to create a variable and wavelength-dependent transmission that can be easily tuned by varying the retardance $R_i$. Based on the Jones calculus, the wavelength dependent transmission $T(R, \lambda)$ can be calculated (for details see Supplementary Information):

$$T(R,\lambda) = \frac{I}{I_0} = \frac{1}{2}[1 + cos(2\pi R_i/\lambda)] \quad (1)$$

which only depends on $R_i$ (and the $\lambda$). The retardance is adjustable by changing the term $R = d\rho(\lambda)$. Here, $d$ represents the thickness of the retarder and $\rho(\lambda)$ is the birefringence of the material. By modulating one of these parameters, the cosine transmission behavior can be spectrally shifted. Therefore, for a specific selected range, the cosine can be transformed into a sine-like shape. In addition, to calculate the spectral phasor, a normalization is needed. This normalization is achieved by setting the retardance term $d\rho$ to 0, for which transmission does not depend on the wavelength, resulting in $T = 1$ over the whole spectral range.

The sine-/cosine-shaped transmission filter can be set to cover any wavelength range by changing the retardance of the variable retarder. This way the wavelength range can be shifted or compressed and stretched (Supplementary Fig. S1), which enables a spectral zoom-in or zoom-out. However, when using this approach, there is a slight error in the calculation of the phasor point. The transmission doesn't follow a perfect cosine wave. Instead, the variable retarder is generating a $cos(1/\lambda)$ function. This results in a shift of the phasor points depending on the magnitude of the error with respect to the perfect sine/cosine function (Supplementary Fig. S1). By tuning the retardance the error can be minimized for the desired spectral region. If the phasor would be calculated in terms of frequency instead of wavelength, a perfect cosine wave would be achieved. Another benefit is the low-cost (starting at 2000 €) and availability of its standard components (Fig. 1c) compared to pricy custom-built sine/cosine interference filters (Supplementary Table S1). We chose a liquid crystal variable retarder (LCVR) as the main element because retardance can be easily adjusted by applying a certain AC voltage (Supplementary Fig. S2a,b) but other optical retarders such as Soleil-Babinet compensators could be used as well. The LCVR can be electronically switched from sine to cosine to no spectral filtering, which eliminates the need for mechanically switching filters.

The workflow for acquiring a hyperspectral image is always the same (Fig. 2a-c). For demonstration purposes a broadband NIR-light and a grating of a monochromator setup (with opened slit) was used to create an image of diffracted light (Fig. 2a). To get the spectral phasor, three images are taken with the HyperNIR module in front of a standard InGaAs NIR camera: One with the sine-shaped filtering ($T_{sin}(R_1, \lambda)$), one with a cosine-shaped filtering ($T_{cos}(R_2, \lambda)$), and one with full transmission ($T_{total}(R_3, \lambda)$) for normalization. In each pixel (x,y) the integral of the transformed spectrum is measured. This makes it possible to calculate each phasor point with the coordinate (G,S) for each individual pixel (details in Supplementary section 1)[36]:

$$S(x,y) = 2 \frac{\int_{\lambda_{min}}^{\lambda_{max}} I(\lambda, x, y) \cdot T_{sin}(R_1, \lambda) \, d\lambda}{\int_{\lambda_{min}}^{\lambda_{max}} I(\lambda, x, y) \cdot T_{total}(R_3, \lambda) \, d\lambda} - 1 = 2 \frac{I_S(x,y)}{I_{total}(x,y)} - 1 \qquad (2)$$

$$G(x,y) = 2 \frac{\int_{\lambda_{min}}^{\lambda_{max}} I(\lambda, x, y) \cdot T_{cos}(R_2, \lambda) \, d\lambda}{\int_{\lambda_{min}}^{\lambda_{max}} I(\lambda, x, y) \cdot T_{total}(R_3, \lambda) \, d\lambda} - 1 = 2 \frac{I_G(x,y)}{I_{total}(x,y)} - 1 \qquad (3)$$

Note, that the 2 and 1 terms are necessary because of equation (1) to normalize G and S and map the phasor onto a unity circle (values between -1 and 1). In summary, the phasor is represented as a scatter plot in which pixels with the same spectrum of light overlay (Fig. 2b). Depending on the point's position, a color can be assigned. As a result, the plain intensity images are transformed into a pseudo-color hyperspectral image (Fig. 2c).

**Tunable phasor-based spectral information**

To characterize the image-based phasor for different wavelengths a monochromator with a xenon-arc lamp was used. Its output was randomly polarized light, with a gaussian spectrum and a full width at half maximum (FWHM) of 15 ± 2 nm. The tunable retardance allows to create different spectral ranges (Fig. 2d-f) e.g. from 900 to 950 nm, 900 to 1200 nm or 900 to 1500 nm. To obtain the best fit to a perfect sine and cosine wave in the specific wavelength regions, the spectral transmissions depending on the LCVR voltage were measured beforehand with a spectrophotometer. The transmission was normalized to the transmission performance of the optical components, which also included the 50% loss of intensity after the first polarizer, due to the random input polarization (Supplementary Fig. S2c). For wavelength ranges narrower than 100 nm, the transmission fit became poorer because of the restricted retardance range of the used LCVR (max R = 11 μm). In the related phasor plots below the transmission curves, it is possible to differentiate clusters. When the spectral range was decreased, the spectral resolution increased. The monochromatic 5 nm steps (Fig. 2d) were equally well separated in the phasor plot as those of 20 nm (Fig. 2e) or 50 nm (Fig. 2f). The spectra with peaks at 1045nm, 1050 nm and 155 nm were indistinguishable in the range 900 to 1200 nm (Supplementary Fig. S3). By zooming into a phasor range of 100 nm and 50 nm, the phasors of the spectra were decomposed. For the range 900 to 1200 nm, the same measurement was repeated with a custom-built interference filter set, which had a nearly perfect match to the sine and cosine function in this range (Supplementary Fig. S4). As expected, in this case the clusters were laying on the edge of the phasor circle and were uniformly distributed. The tunable polarization approach therefore has the tradeoff of a certain difference/error to the perfect phasor position. To get the true phasor position, it would be possible to correct it by creating a look-up table for the G and S values or calibration with NIR standards.

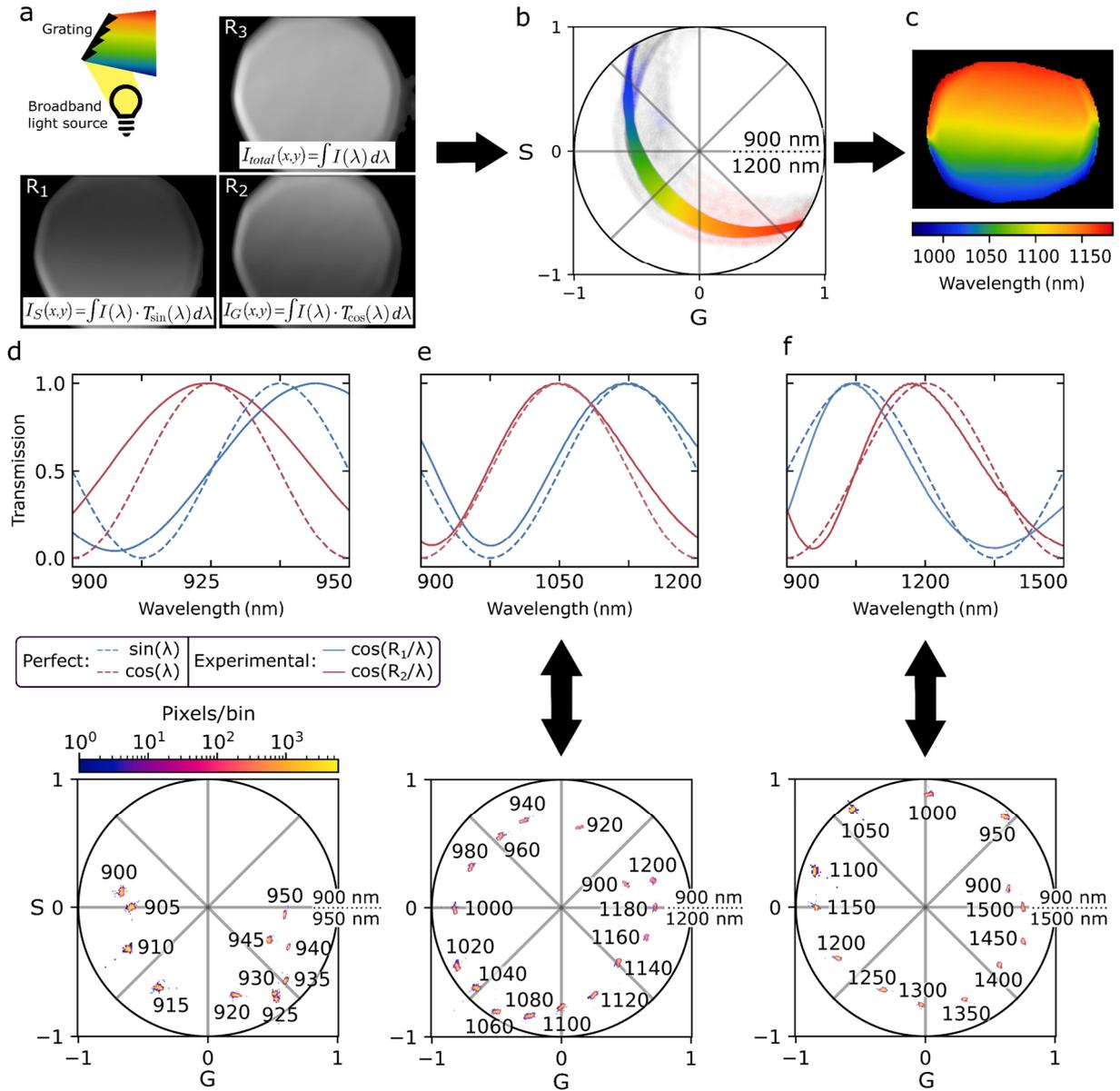

*Fig. 2 Phasor-based hyperspectral imaging with polarization optics. a)-c) Workflow of hyperspectral imaging. a) Model system based on diffraction of a (NIR) white light source from a grating (1200 l/mm). Three images are collected at different retardances ($R_1$, $R_2$, $R_3$), which results in a cosine image ($I_G(x,y)$), a sine image ($I_S(x,y)$) and a total intensity image ($I_{total}(x,y)$). b) Every pixel is transformed into a point in the 2D phasor space. The position of the pixels in the phasor space contains spectral information and can be assigned to a colormap. c) Final HyperNIR image in which the colormap represents the angular position in the phasor plot according to their color in (b). d) – f) Calibration measurements with a monochromatic light source (FWHM = 15 ± 2 nm) for different spectral ranges: d) 900 to 950 nm, e) 900 to 1200 nm, f) 900 to 1500 nm. Upper row: Fit of the experimentally measured transmission and a perfect sine/cosine (dashed line). Lower row: Corresponding phasor plots. Input peak wavelengths of the monochromator are shown in nanometer.*

**Hyperspectral imaging of NIR fluorescence signals**

To demonstrate the (fluorescence) imaging capabilities the HyperNIR module was placed in front of the camera port of a standard wide field microscope. As fluorophores we chose semiconducting single walled carbon nanotubes (SWCNT), which exhibit size-dependent fluorescence between 870 and 2400 nm[39]. They can be conceptualized as rolled-up single layer of graphene. Their structure is described by their roll-up vector/chirality (n,m) and determines their absorption and emission spectrum, which makes them ideal NIR model systems and standards. We selected five of these SWCNTs: (8,3), (6,5), (7,5), (9,4) & (8,4), diluted in 1% sodium deoxycholate (DOC) in water (absorption spectra, see Supplementary Fig. S5a) with emission peaks in the range from 900 to 1200 nm (Fig. 3a). The (8,3)-, (6,5)- and (7,5)-SWCNTs showed distinct phasor clusters, which correspond to their emission spectra (Fig. 3b). Due to the slightly broader emission of the (6,5)-SWCNTs compared to the (8,3)-SWCNTs, the pixels were closer to the center of the phasor plot. For the 900 nm - 1500 nm spectral phasor range (9,4)- and (8,4)-SWCNTs were not distinguishable. Thus, a spectral zoom-in into the range 900 to 1200 nm was performed (Fig. 3c) by adjusting the spectral range of HyperNIR. Consequently, the phasor clusters of (9,4)- and (8,4)-SWCNTs separated despite having nearly identical fluorescence spectra with center wavelengths separated by only 5 nm. By reducing the phasor range from 600 nm to 240 nm the distance between the mean values in the phasor plot could be increased by 100 % (Supplementary Fig. S6). These results show the advantage of using a tunable filter instead of a fixed hardware filter.

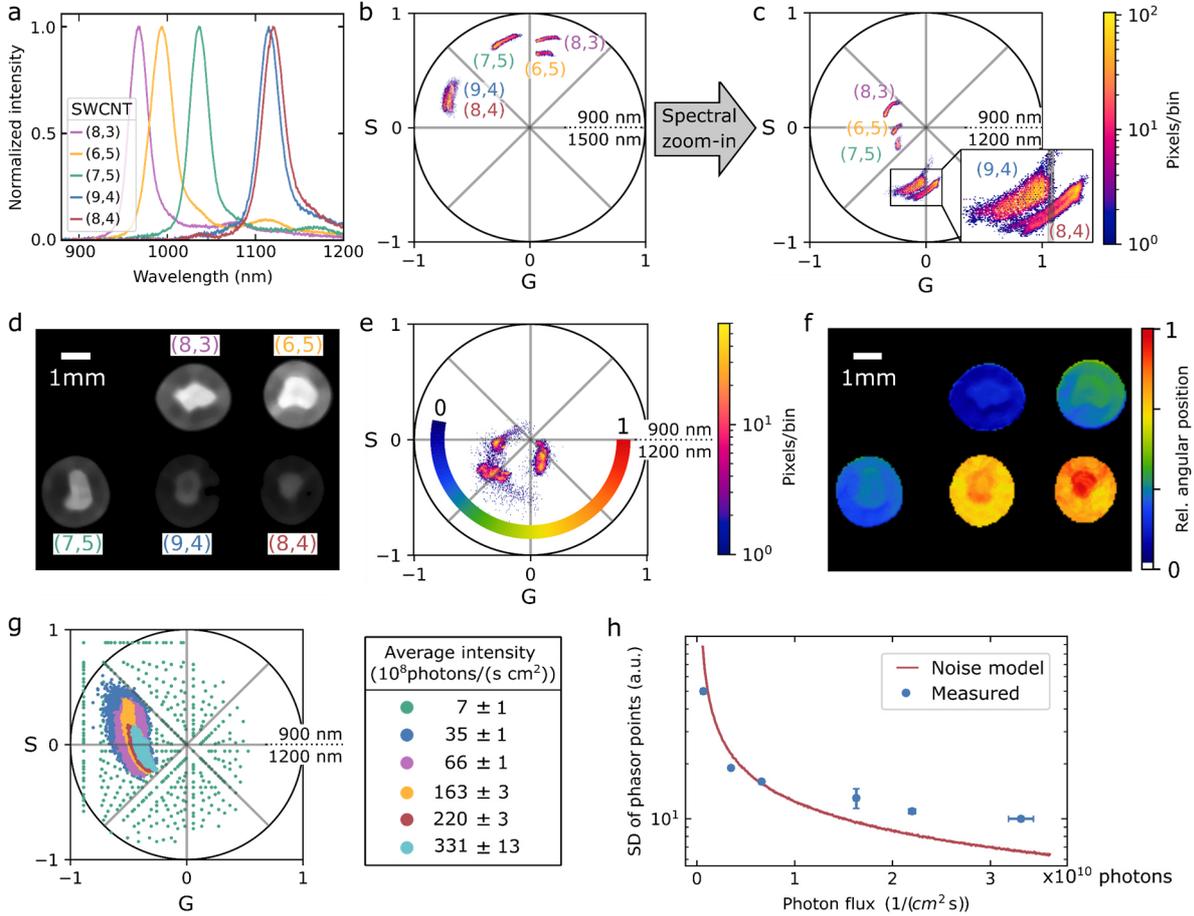

*Fig. 3: Hyperspectral imaging of NIR fluorescence signals.* a) NIR fluorescence spectra of SWCNTs of different chiralities (n,m). b) Phasor plot of the SWCNTs from (a) in a homogenous solution in a broad wavelength range from 900 to 1600 nm. c) Spectral zoom-in to the phasor range 900 to 1200 nm (by adjusting the retardance of HyperNIR) allows to differentiate (9,4)- and (8,4)-SWCNTs. d) NIR fluorescence image of SWCNTs drop-casted in small spots on a polymer membrane. e) HyperNIR transformation into the spectral phasor space. h) Based on the angular position in the phasor plot in (e), the pixels are color-coded. g) Phasor plot for different levels of photon flux. h) Phasor uncertainty decreases with average light intensity (blue). The red line corresponds to a camera noise model (EMVA Standard 1288, see methods for details)[40]. Data = Mean ± SD, n = 3.

The five SWCNT-chiralities were also dropcasted on a polymer membrane (Supplementary Fig. S5b). The array was imaged with the same microscope using a lens (f = 150 mm, de-magnification 0.75x) in place of the objective (Fig. 3d). In contrast to the phasor plot of the solutions, the absolute position of the phasor was shifted for each chirality (Fig. 3e). A reason for this could be the background signal of the membrane, which was subtracted by measuring the phasor of the membrane. We then created a color-coded hyperspectral image based on the angular position of the pixels in the phasor (Fig. 3f). Additionally, spectral differences in single spots were visible that are most-likely caused by inhomogeneous drying of SWCNTs and aggregation, which is known to red-shift/broaden SWCNT spectra[39]. The SWCNT emission can be tuned by introducing guanine quantum defects (from 990 nm to 1040 nm), which results in a peak shift and broadening[41] (Supplementary Fig. S7a). Such changes can be monitored with HyperNIR and the composition of samples with SWCNTs with and without defects could be determined based on the phasor position (Supplementary Fig. S7b, Table S2).

To understand the limits of HyperNIR at low light intensities, a solution of (6,5)-SWCNTs was imaged at six different concentrations (Fig. 3g). The smaller the average intensity became, the greater was the spread of the phasor points. For an average intensity of about $10^9$ photons/(cm² s), the noise of the InGaAs camera showed a broad distribution in the phasor. Compared to the other five data points, there was no cluster observable. We then calculated a signal to noise ratio from the sum of the S and G standard deviations for the G and S values and plotted it against the average intensity in the image (Fig. 3h). It is well fitted by a noise model for cameras based on the "EMVA Standard 1288"[40] (see Methods section). These results show that there is a photon limit of ~$10^9$ photons/(s cm²) to accurately determine the phasor of an image. It reflects the uncertainty one should expect for wavelength measurements of signals with low photon statistics. The fluorescence quantum yields of NIR fluorophores are typically low[42] and therefore this is more relevant in low photon number fluorescence experiments than NIR reflection/absorption measurements with many photons.

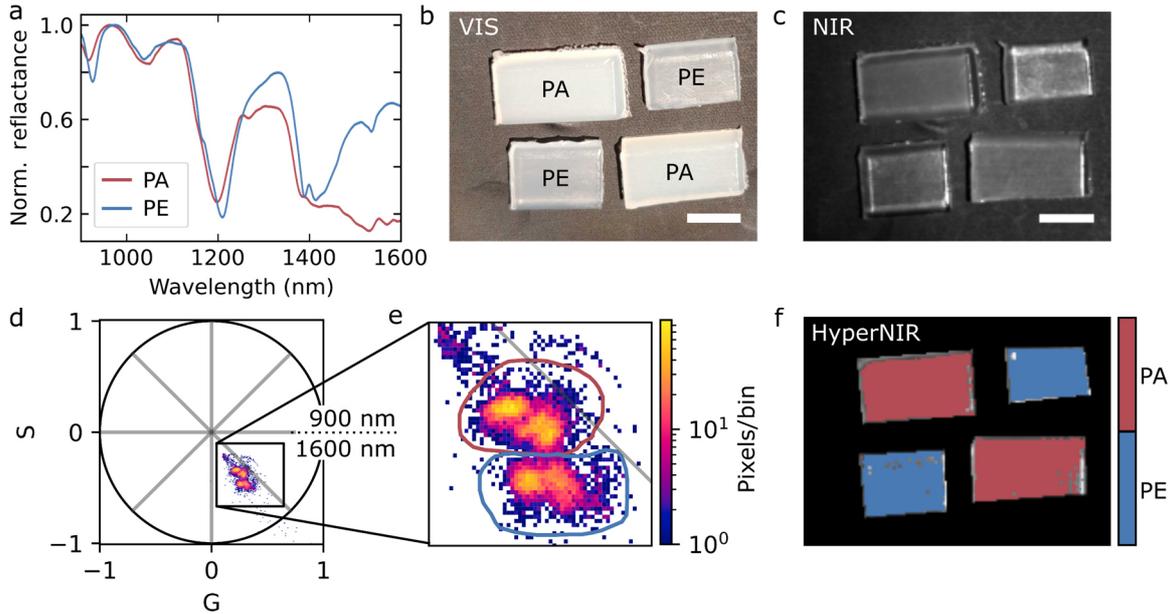

**Fig. 4: Hyperspectral NIR discrimination of plastics.** a) NIR reflectance spectra of the plastic polymers polyethylene (PE) and polyamide (PA). b) Visible image of pieces of PA and PE. c) Corresponding NIR intensity image. d) Corresponding HyperNIR phasor plot for the range 900 to 1600 nm. e) Magnified plot of the phasor points in (d). f) HyperNIR image (dual-color coding based on the pixel position in the phasor plot according to the blue and red circles in (e). Scale bar = 0.5 mm, exposure times = 15 ms.

**Macroscopic HyperNIR reflectance imaging**

Next, we tested HyperNIR reflectance imaging of macroscopic samples. The samples were illuminated by a tungsten-halogen lamp at an approximate angle of 70 degrees (Supplementary Fig. S8a). The diffuse reflected light was then imaged through two demagnifying lenses (M = 0.15x). As a homogeneous reference, a diffuse reflectance standard with a near perfect Lambertian reflection was utilized. To cover the full spectral reflection of the NIR, the phasor range was chosen from 900 to 1600 nm. At first the reference was measured for each of the three retardance settings. We observed certain inhomogeneous intensity patterns (Supplementary Fig. S8b) caused by an imperfect illumination and aberrations. By calculating a generic correction matrix for the G and S components, a dependence on the LCVR setting became visible (Supplementary Fig. S8c). This inhomogeneity is well known in liquid crystal variable retarders[43,44], but by applying the correction matrices this effect is suppressed.

The identification and differentiation of plastics is highly relevant in waste management and recycling applications[45]. To demonstrate the capabilities of HyperNIR, we imaged two different plastics, namely polyethylene (PE) and polyamide (PA), next to each other. The reflectance spectra were very similar besides differences in the region from 1400 to 1600 nm (Fig.4a). In the visible spectral range, the color of the pieces was nearly the same (Fig. 4b), which was also found for the monochrome NIR image (Fig. 4c). For a better classification we calculated the phasor for each pixel (Fig. 4d) and observed two separate clusters (Fig.4e). We then applied a polygon mask to color code the pixels in the image accordingly (Fig. 4e). We also imaged with multi-order sine and cosine filters (full wave for the range 900 to 1200 nm, 1100 to 1600 nm, Supplementary Fig. S9a). In this case the phasor is not unambiguous but between 900 and 1200 nm the reflectance spectra showed no significant spectral differences. Therefore, this range did not contribute to a difference in the phasor as much as the 1100 nm to 1600 nm range, for which distinct peaks and an overall difference in the spectra was observed. By applying this filtering, we again found two clusters in the phasor space, which was used for binary color coding (Supplementary Fig. S9b-f). This example shows that even without additional bandpass filters it is possible to use different spectral ranges of HyperNIR. Overall, we show that HyperNIR enables fast discrimination of plastic polymers from 2D images, which is an advantage compared to using 1D hyperspectral cameras in combination with the necessity of moving the sample[46].

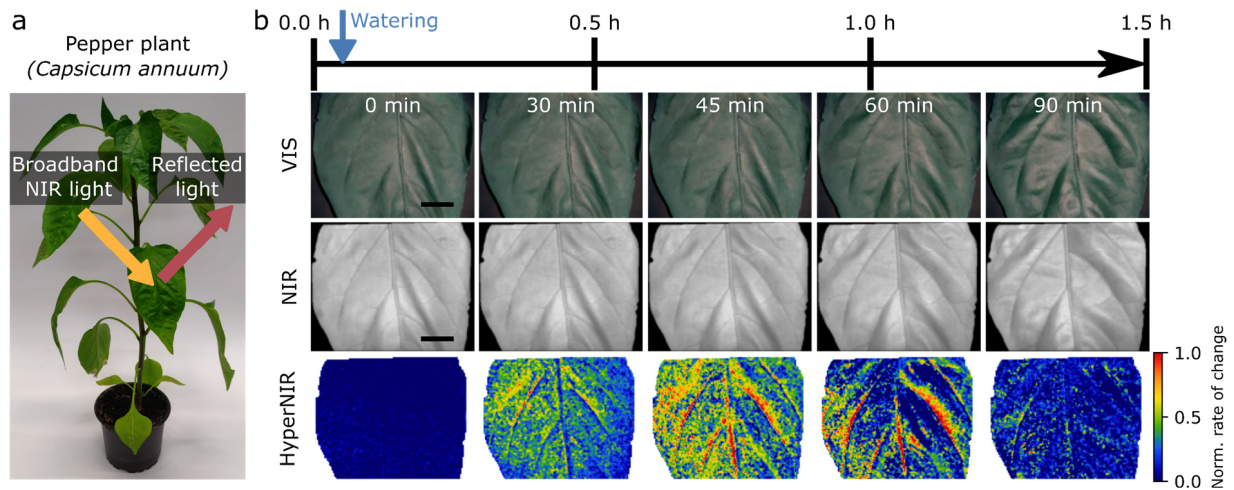

**Fig. 5: NIR hyperspectral imaging of leaf water uptake.** *a) Pepper plant (capsicum annuum) and design of label-free reflectance imaging. b) The RGB camera (VIS), NIR-monochrome (NIR) and HyperNIR image from a leaf of a dry plant were taken over 1.5 h with a hyperspectral frame rate of 0.2 /s. After approx. 1 min the plant was watered. In the HyperNIR images the normalized rate of change is visualized, which represents the movement of the phasor relative to the phasor position at the beginning. It represents leaf water uptake based on the water combination modes in the NIR. Scale bar = 1 mm, exposure time (NIR & HyperNIR) =7 ms.*

Next, we wanted to explore the potential of HyperNIR for imaging of biological systems. Tools that monitor plant stress are necessary for precision agriculture[47]. Water is essential for plants and has prominent features in the NIR. Therefore, we chose to monitor water uptake in plants. Plant leaves show high reflection in the range from 900 to 1350 nm, which corresponds to light scattering at the spongy cell structure of the leaves[48]. At around 1450 nm the water absorption peak dominates the spectrum (Supplementary Fig. S10). By using this feature in a phasor, we wanted to measure the spatial and temporal distribution of the water uptake inside a leaf. As a model system, we used pepper plants (*Capsicum annuum*, Fig. 5a), which were not watered for 3 days. One leaf of the dry plant was placed in the field-of-view and imaged with a framerate of 0.2 hyperspectral cubes per second. This framerate is limited by the switching times of the LCVR between the three retardance settings (Supplementary Fig. S11). Then the relative change of the phasor (Euclidian distance) relative to the start of the measurement was calculated for each time point and color-coded (Supplementary Fig. S13, Fig. 5b). The plant was watered 1 min after the video started, but the first 15 min showed nearly no change (only random fluctuations) in the HyperNIR image (Supplementary Video 1, Supplementary Fig. S13). 25 min after watering water transport in the capillaries became visible and after around 30 – 60 min also the surrounding tissue showed changes in the phasor indicating an increase of the water concentration in the tissue. Such patterns were not observed in control experiments without water exposure (Supplementary Fig. 12). After 1 h the rate of change decreased, resulting in a nearly homogenous image (see Supplementary Fig. S14 for other leaves/plants). These results show that HyperNIR enables *in vivo* imaging of plant health without the need of dyes, and this could be extended to biochemical compounds that are linked to other health states such as lipids. Moreover, one could use this approach to monitor processes related to photosynthesis and assess plant productivity, which is highly interesting for precision agriculture.

**Discussion**

HyperNIR is a powerful contactless technique for hyperspectral NIR fluorescence or label-free (reflectance) imaging. It uses mapping of spectral information into a 2D space (phasor) by imaging through different filters similar to a hardware-based Fourier transformation. We demonstrate a tunable and straightforward optical approach to obtain the spectral phasor using polarization optics. By adjusting the retardance of the LCVR, we were able to shift the phasor range to distinguish different NIR fluorophores. It opens the possibility to perform a spectral zoom in and zoom out. This could be also useful to avoid bandpass filters in situations when the multi-order cosine and sine filters are not unambiguous. In videos we achieved so far frame rates of 0.2 hyperspectral cubes per second. The software calculation of the spectral image takes less than 5 ms on a standard laptop computer, which demonstrates the potential for live imaging. The limiting factor for acquisition speed is the switching time of the LCVR. We used a commercially available one (Thorlabs) but faster models would directly translate into higher frame rates. A way to overcome current limits could be to split the LCVR in a few cells and drive every cell separately[49]. Another approach is advanced engineering of LCVR cell by using polymer-stabilized or polymer-dispersed liquid crystals, which can decrease the switching times[50,51,52]. The maximum frame rate one can achieve with HyperNIR is 3 times the frame rate for monochromic images because one requires 3 images for the phasor calculation. Moreover, the used optical elements are readily accessible and can be bought starting at a price of 2000 € (Supplementary Table S1). Therefore, HyperNIR can be seen as a straightforward add-on to equip normal cameras with very fast hyperspectral imaging capabilities.

The HyperNIR approach is also directly applicable to the Vis range of the spectrum if one uses the appropriate optical components. However, the NIR contains more relevant spectral information for label-free fingerprinting. Additionally, the spectral Vis range is smaller and typically Si-based Vis cameras offer a much higher resolution (number of pixels). Therefore, the NIR is the spectral window in which HyperNIR can use and leverage its advantages. Another option is to use Vis-SWIR cameras to further increase the spectral range and the information content of the images.

As we showed HyperNIR can be used for fluorescence imaging as well as for label-free imaging. Fluorescence imaging in this range is especially useful in the biomedical field and spectral multiplexing is desired to increase the information content or observe wavelength shifts[26]. NIR spectroscopy is often used as fast method to identify materials, and our approach could make the imaging counterpart available for a much bigger community. Moreover, one can anticipate that using 2D chemometrics and machine learning approaches allow tailored analysis of HyperNIR images and videos in the future.

Overall, we present a powerful method to obtain hyperspectral information by attaching a small footprint HyperNIR module to a camera. This could be especially relevant for experiments and applications in which a low mass and volume is desired such as drones. In conclusion, we anticipate great potential for fast NIR hyperspectral imaging. It can pave the way for imaging of multiple NIR fluorophores or for label-free imaging for environmental monitoring and precision agriculture.

## Methods

### HyperNIR module & acquisition

The HyperNIR module consists of two linear polarizers (LPNIR100, Thorlabs) and a liquid crystal variable retarder (LCC1115-B, Thorlabs), which was operated by a LC-controller (KLC101, Thorlabs). For all measurements the frequency of the LCVR was kept at 5 kHz. For acquisition, the LCVR switching process was automated via the python SDK. As NIR camera an InGaAs camera (Xeva 1.7 320 TE3 USB 100, Xenics) with an image format of 256x312 pixels was used. For the acquisition the LCVR was controlled by the python SDK (K-Cube LC Controller Software, Thorlabs) and the camera was operated with "µManager" and the python package *pycromanager*. The G and S values were calculated in python with regards to equations (2) and (3) with minor adjustments (see Supplementary section 2). The phasor plots were created using a 2D-histogram (*matplotlib.pyplot.hist2d*) with bin sizes adjusted to the amount of analyzed pixels.

### Transmission measurement & LCVR voltage settings

Transmission spectra were acquired with a spectrophotometer V-780 (JASCO Deutschland GmbH). With a customized 3D printed adapter, the 30 mm cage system of HyperNIR was centered in the light path. The transmission was measured for the wavelength range 750 to 1600 nm. The measured curves were normalized by the transmission performance of the two polarizers and the LCVR (Supplementary Fig. S2c). The applied voltage at the LCVR was changed from 0 to 4 V in 0.005 V steps and subsequently the transmission spectrum was recorded. The stack of measured transmission curves was correlated with a perfect sine and cosine function for a defined spectral phasor range. The best fit was calculated by a python script using the Pearson correlation. The curves with the highest R-value were then used for the phasor acquisition. For the best fit, both negative and positive cosine and sine functions were evaluated. If a negated cosine or sine represents the best fit, this was considered in the phasor transformation (Supplementary section 2).

**Characterization with monochromatic light setup**

The monochromatic source was powered by a xenon-arc lamp (300 W) and split into spectral bands by a monochromator (MSH-150, LOT-Quantum Design GmbH, f/4.6, f = 150 mm) using a diffraction grating (MSG-T-1200-500, 30 mm × 30 mm, 1200 l/mm). The light was then passed through the polarization optics and was collected by a liquid light guide (Ø5 mm core, LLG5-4Z, Thorlabs), which was coupled out by a collimator (SLSLLG1, Thorlabs) into an Olympus IX73 inverted microscope (Objective: Olympus MPlan N, ×5/NA 0.10). The tube lens of the microscope focused the light onto the InGaAs camera. The image of the diffracted light was post-processed by subtracting a background signal and setting a threshold. To suppress pixel errors, the image was filtered with a median filter (*scipy.signal.medfilt2d*, filter window = (3,3)). Pixels with phasor points laying outside of the unity circle were ignored for the color-coded image. Based on the phasor points angle, a colormap was applied. For the monochromatic signals the same post-processing was done.

**NIR fluorophore preparation**

Single-walled carbon nanotubes (SWCNT) were used as versatile NIR model fluorophores. As previously reported[53,54], the separation of chiralities (9,4), (7,5), (8,4), (8,3), and (6,5) single-walled carbon nanotubes (SWCNTs) was achieved using an aqueous two-phase extraction (ATPE) method. Initially, a raw (6,5) chirality enriched SWCNT mixture (CoMoCat, Sigma-Aldrich) was suspended in a solution containing dextran (MW 70,000 Da, TCI) and polyethylene glycol (PEG, MW 6,000 Da, Sigma-Aldrich) to create two distinct aqueous phases. A fixed concentration of sodium deoxycholate (DOC, 0.05% m/v, Sigma-Aldrich) and gradually increasing concentrations of sodium dodecyl sulfate (SDS, 0.5% - 1.5% m/v, Sigma-Aldrich) were employed. Subsequently, a semiconducting-metallic sorting step removed metallic SWCNTs from the chiral-enriched fractions[53]. In this step, sodium cholate (SC) was added, and the overall surfactant concentrations were adjusted to 0.9% SC, 1% SDS, and less than 0.02% DOC, followed by the addition of sodium hypochlorite (NaClO, 10–15% available chlorine, Honeywell) at 5 μL/mL of a solution prediluted to a 1/100th concentration in water as the oxidant. All sorted species were reconcentrated and adjusted to 1% DOC (m/v) using iterative concentration and dilution cycles in a pressurized ultrafiltration stirred cell (Millipore) equipped with a 300 kDa cutoff membrane. The microscopic measurements were done by pipetting 150 μL of each SWCNTs solution in a black 96 microtiter well plate with clear bottom (Fisher Scientific). The SWCNT array was prepared on a polyvinylidene fluoride (PVDF) membrane, which has a low background fluorescence in the visible and NIR (Immobilon®-FL PVDF-Membran, Milipore). For every SWCNT chirality, 1.5 μL were drop-casted on the membrane and dried on a hot plate operated at 60°C.

For the concentration-dependent phasor, (6,5)-enriched SWCNTs without purification were used. Compared to the chirality-pure SWCNTs, these SWCNTs were solubilized by biofunctionalization with the ssDNA sequence $(GT)_{15}$. In short, $(GT)_{15}$ ssDNA (200 μL, 100 μM in PBS, Sigma Aldrich) was mixed with SWCNTs (100 μL, 2 mg/mL in PBS) and tip-sonicated (Fisher Scientific Model 120 Sonic Dismembrator, 35% amplitude, 20 min pulsed: 9 s on, 1 s off). The resulting suspension was centrifuged (2x, 21000 g, 30 min) and the supernatant was transferred into a different reaction vessel. The SWCNTs were prepared in phosphate buffered saline (1xPBS) in the concentrations 0.1 nM, 0.5 nM, 1 nM, 3 nM, 5 nM and 10 nM[55]. For the measurement, triplicates of 200 μl of each concentration were pipetted into a 96 microtiter well plate. Additionally, for background subtraction triplicates of 200 μL 1xPBS were used.

The preparation of SWCNTs with guanine quantum defects ($G^d$) followed a previously published protocol[41]. For the covalent functionalization of guanine containing ssDNA with the SWCNTs, the $(GT)_{15}$-functionalized SWCNTs (189 μL, 10.6 nM in PBS) were mixed with Rose Bengal (10.1 μL, 330 μM in $H_2O$, Sigma-Aldrich) in a 96-microwell plate and irradiated (Analytical Sales Lumidox 96-Well Green LED Array + LED Controller, 25 mA, ≈ 527 nm). After 10 min, air was bubbled through the mixtures and the respective wells were irradiated for an additional 5 min. Free DNA and Rose Bengal were removed via dialysis (Spectrum Laboratories SpectraPor membranes, molecular weight cut-off: 300 kDa) for five days with buffer changes twice a day. The suspensions were briefly tip-sonicated for redispersion (35 % amplitude, 20 s), centrifuged (21000 g, 20 min) and the supernatant was used for further studies. The $G^d$-SWCNTs were mixed with chirality pure (6,5)-SWCNTs, which both were prepared in the concentration 1 nM. Mixtures of (6,5)-SWCNTs without and with defects were prepared in ratios of 50:50, 75:25, 90:10, 95:5 in a total volume of 200 μL. To gain the phasor of the pure samples 200 μL of each were used a reference.

**NIR fluorophore spectroscopy and imaging**

Emission spectra of SWCNTs were acquired with a custom-build microscope setup, which was described in previously studies[41], equipped with a 561 nm Laser (gem-561, Novanta Photonics) and a InGaAs spectrograph (Shamrock 193i + Andor iDus InGaAs 491 detector, Andor Technology). Background signals were corrected in the spectrometer software using a spectrum of the buffer.

All SWCNT measurements were performed at the inverted microscope (Nikon Eclipse Ti2). It consisted of an excitation light (CoolLED pE300 Lite, 100% power), which was transmitted through a bandpass (for (6,5)-SWCNTs: 560 ± 40 nm (F47-561, AHF Analysentechnik) or for measurement with more SWCNT chiralities: 495 ± 130 nm (FSR-BG40, Newport)) and a 804 nm longpass dichroic mirror (AHF Analysentechnik F38-801) before being focused by an objective (CFI Plan Fluor DL 10XF, Nikon). Emission light passed through a 840 nm longpass filter (F47-841, AHF Analysentechnik). The HyperNIR setup and Xeva camera were positioned at the output of the left camera port.

For all solution measurements: The focus was adjusted to approximately the middle of the sample in order to achieve maximal signal and a homogenous distribution. The exposure times for the images were set to 3 s. A background signal of the excitation source (LED) and dark noise was subtracted from the fluorescence image for each sine, cosine and total image. Due to strong circular vignetting at the edges, a circular mask was applied. The images were filtered with a median filter (filter window = (3,3)). For the concentration-dependent phasor analysis, the spreading of the phasor clusters was evaluated by calculating the sum of the standard deviation of the G and S components over all phasor points for each concentration. For spectral unmixing of SWCNTs with and without guanine quantum defects the fractions were calculated according to Supplementary section 3.

For imaging the SWCNT array a 150 mm lens (AC254-150-AB-ML, Thorlabs) instead of a microscope objective was used. The remaining optical setup was the same as for the SWCNT solution imaging. As a background signal a membrane without any SWCNTs were imaged. Due to the inhomogeneous drying of the samples, a stronger median filter (filter window = (9,9)) was applied.

**Noise model**

The used noise model is based on the "EMVA Standard 1288: Standard for Characterization of Image Sensors and Cameras"[40]. The overall variance $\sigma_y^2$ of the detected signal at the camera is the linear sum of all noise sources

$$\sigma_y^2 = K^2 \sigma_d^2 + \sigma_q^2 + K(\mu_y - \mu_{y,dark}) \tag{4}$$

where $K$ is the system gain of the camera, $\sigma_d^2$ represents the read-out noise and $\sigma_q^2$ stands for the quantization noise. The noise is mainly influenced by the amount of signal $\mu_y$ reaching the detector, which is subtracted by the dark signal $\mu_{y,dark}$. For the used InGaAs camera the parameters can be found in the data sheet: $\sigma_d^2 = 1000\ e^-$, $\sigma_q^2 = \frac{1}{12} DN^2$ (12-bit), $K = 1.1\ 10^{-3} DN/e^-$. The dark signal $\mu_{y,dark}$ was neglected in the theoretical calculation since the measured intensities were corrected by the background and the dark signal. For the noise calculation, images with 256x312 pixels with a normal gaussian random distribution (Python function *numpy.random.normal*) were generated. This was done for a signal $\mu_y$ in the range from 10 to 600 DN. For the sine and cosine image a fraction of this signal intensity was used, and the fraction was calculated by integrating the transmitted sine and cosine part of (6,5)-SWCNT spectra over all wavelengths and normalized to the total intensity. For a realistic intensity profile, a circular gaussian mask was applied. The photon flux on the camera was calculated by

$$\Phi = \frac{\mu_y}{K\ QE\ A\ t_{exp}} \left[\frac{\#\ photons}{cm^2\ s}\right] \tag{5}$$

where $QE = 0.56$ (at 990 nm) is the quantum efficiency, $A = 9 \cdot 10^{-6}\ cm^2$ the pixel area of the camera and $t_{exp} = 3\ s$ the exposure time of the camera.

**Macroscopic reflectance imaging**

For HyperNIR imaging of macroscopic samples, a custom-built setup was used. The camera was mounted vertically on an aluminum strut profile. The polarization optics were placed in between a magnifying two lens system. On the object-side a 200 mm plano-convex lens (LA1708, Thorlabs) was used, whereas the image was generated by a 30 mm achromatic lens (AC254-030-AB, Thorlabs). This resulted in a magnification of 0.15x for the overall system. Based on this, a field of view of 51 mm x 64 mm was obtained. Before the imaging lens an iris diaphragm (SM1D25, Thorlabs) was a placed to adjust the incident light and improve the overall image quality. To filter out the visible spectral range, a 840 nm longpass filter (AHF Analysentechnik F47-841) was positioned between the object lens and the first polarizer. As a NIR illumination source a quartz tungsten-halogen lamp (QTH10/M, Thorlabs) illuminated the sample at an approximate angle of 70°. For imaging we used short exposure times between 5 and 50 ms. A correction matrix for G and S (Supplementary Fig. 8b and section 3) were obtained by measuring a white diffuse reflectance standard (Spectralon Ø 31.75 mm, Labsphere). Therefore, the reflectance standard was moved to all angles of the field-of-view and the images were then mapped together. To suppress reflections from the background we used a matte black aluminum foil (BKF12, Thorlabs) as a base for the samples.

The plastic pieces were purchased from PlasticsEurope Deutschland e.V. as a sample collection. The samples were pure polyamide and pure polyethylene pieces with a thickness of 2.5 mm. In the post-imaging analysis, we cut out the four plastic pieces and set the rest of the background to 0. The G and S values were corrected with a correction matrix obtained by a white reflectance standard (Supplementary Fig. 8b and section 3). Polygon masks were used to define clusters in the phasor plot and create binary colormaps.

For the leaf water uptake experiments, a pepper plant (*Capsicum annuum*) was used. It was bought as a seedling in a pot at a local gardening store. To measure the water uptake the plant was not watered for three to five days. One leaf was placed in the camera view and imaged with a hyperspectral frame rate of 0.2 /s (Acquisition cycle: set 1.94 V → pause 1.5 s → set 2.19 V → pause 0.5 s → set 25 V → pause 3 s → set 1.94 V, see Supplementary Fig. S11). After imaging the dry leaf for one hour the plant was watered with a volume of approx. 40 mL. After that, the video continued for an additional 1.5 hour. A polygon mask was used to extract the parts of the image that contained a leaf, the rest of the image was set to 0. An average phasor point for each pixel as a reference for a dry leaf was calculated over the hyperspectral frames of the first 5 min. These phasor points were used as reference position for the upcoming frames. For each pixel the movement of the phasor (Euclidean distance) in contrast to the reference position was calculated. Based on this, the rate of change was calculated by forming the linear regression (least squares) over a range of 10 min. Afterwards all values were normalized to the rate of change which was present at 99% of the cumulative histogram (calculated over all frames and pixels). All G and S values were again corrected by the correction matrices. The Vis images were acquired by a webcam (Logitech) with 1920x1080 pixel.

**Acknowledgments**

This work was supported by the Fraunhofer Internal Programs under Grant No. Attract 038-610097. Funded by the Deutsche Forschungsgemeinschaft (DFG, German Research Foundation) under Germany's Excellence Strategy – EXC 2033 – 390677874 – RESOLV. This work was supported by the "Center for Solvation Science ZEMOS" funded by the German Federal Ministry of Education and Research BMBF and by the Ministry of Culture and Research of North Rhine-Westphalia. This work was funded by the VW Foundation.

**Conflict of Interest**

S.K. and J.S. are listed as inventors on a pending patent application describing the HyperNIR module presented in this study.

# Supplementary Information

**Hyperspectral near infrared imaging using a tunable spectral phasor**

Jan Stegemann[1,2], Franziska Gröniger[2], Krisztian Neutsch[1], Han Li[3,4], Benjamin Flavel[5], Justus Tom Metternich[1,2], Luise Erpenbeck[6], Poul Petersen[1], Per Niklas Hedde[7], Sebastian Kruss[1,2*]

[1]Department of Chemistry and Biochemistry, Bochum University, Bochum, Germany

[2]Fraunhofer Institute for Microelectronic Circuits and Systems, Duisburg, Germany

[3]Department of Mechanical and Materials Engineering, University of Turku, FI-20014 Turku, Finland

[4]Turku Collegium for Science, Medicine and Technology, University of Turku, FI-20520 Turku, Finland

[5]Institute of Nanotechnology, Karlsruhe Institute of Technology, Karlsruhe, Germany

[6]Department of Dermatology, University Hospital Münster, Münster, Germany

[7]Beckman Laser Institute & Medical Clinic, University of California, Irvine, CA, USA

Corresponding author: sebastian.kruss@rub.de

# Extended Methods

The phasor itself is the first harmonic of the Fourier transform of a spectrum $I(\lambda)$[1]:

$$\hat{F} = \int_{-\infty}^{\infty} I(\lambda) \cdot e^{-i\frac{2\pi}{L}\lambda} d\lambda \tag{1}$$

where $L = (\lambda_{max} - \lambda_{min})$ defines the analyzed spectral range. As a representation of the phasor, the real and imaginary parts of the Fourier transform G and S are used as coordinates to map the spectrum onto a 2D space[1]:

$$G = \frac{\int_{\lambda_{min}}^{\lambda_{max}} I(\lambda) \cdot \cos\left(\frac{2\pi(\lambda - \lambda_{min})}{(\lambda_{max} - \lambda_{min})}\right) d\lambda}{\int_{\lambda_{min}}^{\lambda_{max}} I(\lambda) d\lambda} \tag{2}$$

$$S = \frac{\int_{\lambda_{min}}^{\lambda_{max}} I(\lambda) \cdot \sin\left(\frac{2\pi(\lambda - \lambda_{min})}{(\lambda_{max} - \lambda_{min})}\right) d\lambda}{\int_{\lambda_{min}}^{\lambda_{max}} I(\lambda) d\lambda} \tag{3}$$

The components G and S are normalized to the entire integrated spectrum, where $\int_{\lambda_{min}}^{\lambda_{max}} I(\lambda) d\lambda$ corresponds to the toal intensity over the full spectral range. A polar diagram is appropriate for visualization of the phasor points. Thus, a vector with an angle $\varphi$ and modulation $M$ can be calculated:

$$\varphi = \tan^{-1}\left(\frac{S}{G}\right) \tag{4}$$

$$M = \sqrt{S^2 + G^2} \tag{5}$$

In Fig. 1b (main manuscript) two exemplary spectra and their representation in a phasor plot are shown. A full circle represents the whole analyzed spectral range. Therefore, the angle reflects the wavelength of the peak of the spectrum. The modulation, on the other hand, corresponds to the broadness of the spectrum. The broader the spectrum, the closer to the center lies the phasor point.

For generating the spectral phasor directly with hardware components, the pixelwise spectrum $I(x, y, \lambda)$ is weighted with a sine and cosine transmission filter for a desired spectral range ($\lambda_{min}, \lambda_{max}$) resulting in an intensity at each pixel of[2]:

$$I_{sin}(x, y) = \int_{\lambda_{min}}^{\lambda_{max}} I(x, y, \lambda) \cdot \sin\left(2\pi \frac{\lambda - \lambda_{min}}{\lambda_{max} - \lambda_{min}}\right) d\lambda \quad \text{(analogous for } I_{cos}\text{ )} \tag{6}$$

Since the filters have a transmission range from 0 to 1, the pixel values need to be shifted to the interval from -1 to +1 to obtain $G$ and $S$ within the unity circle[2]:

$$G(x, y) = 2 \frac{I_{cos}(x, y)}{I_{total}(x, y)} - 1 \tag{7}$$

$$S(x, y) = 2 \frac{I_{sin}(x, y)}{I_{total}(x, y)} - 1 \tag{8}$$

## 2. Theoretical background of the polarization filtering to create a cosine transmission

To generate a cosine shaped transmission, a variable retarder is placed between two linear polarizers. Based on the Jones calculus, the transformation of the polarization state can be calculated. For a linear polarizer operating at an angle $\beta$ the Jones matrix is[3]:

$$M_P = \begin{bmatrix} \cos^2(\beta) & \frac{1}{2}\sin(2\beta) \\ \frac{1}{2}\sin(2\beta) & \sin^2(\beta) \end{bmatrix} \tag{9}$$

and the matrix for a variable retarder is given by[3]:

$$M_R = \begin{bmatrix} 1 & 0 \\ 0 & e^{i\varphi} \end{bmatrix} \quad (10)$$

where $\varphi$ is the set phase retardation of the variable retarder, which is given by $\varphi = 2\pi R/\lambda$ and depends on the wavelength $\lambda$ and the retardance $R$.

When the incoming light is randomly polarized, the light has a linear polarization of angle $\alpha$ after the first linear polarizer[3]:

$$E_{in} = \begin{bmatrix} \cos(\alpha) \\ \sin(\alpha) \end{bmatrix} \quad (11)$$

This linear polarized light is passed through the variable retarder and the second polarizer, which acts like a filter for polarization. At the end, the output polarization vector $E_{out}$ can be calculated as follows:

$$E_{out} = M_P \cdot M_R \cdot E_{in} = \begin{bmatrix} \cos^2(\beta) & \frac{1}{2}\sin(2\beta) \\ \frac{1}{2}\sin(2\beta) & \sin^2(\beta) \end{bmatrix} \cdot \begin{bmatrix} 1 & 0 \\ 0 & e^{i\varphi} \end{bmatrix} \cdot \begin{bmatrix} \cos(\alpha) \\ \sin(\alpha) \end{bmatrix} \quad (12)$$

For operating both linear polarizers at angle $\alpha = \beta = 45°$, the equation can be simplified to:

$$E_{out} = \begin{bmatrix} 1/2 & 1/2 \\ 1/2 & 1/2 \end{bmatrix} \cdot \begin{bmatrix} 1 & 0 \\ 0 & e^{i\varphi} \end{bmatrix} \cdot \begin{bmatrix} \sqrt{2}/2 \\ \sqrt{2}/2 \end{bmatrix} = \frac{\sqrt{2}}{4} \cdot \begin{bmatrix} 1 & 1 \\ 1 & 1 \end{bmatrix} \cdot \begin{bmatrix} 1 & 0 \\ 0 & e^{i\varphi} \end{bmatrix} \cdot \begin{bmatrix} 1 \\ 1 \end{bmatrix} \quad (13)$$

which gives a final output vector for the polarization of:

$$E_{out} = \begin{bmatrix} E_{0x} \\ E_{0y} \end{bmatrix} = \frac{\sqrt{2}}{4} \begin{bmatrix} 1 + e^{i\varphi} \\ 1 + e^{i\varphi} \end{bmatrix} \quad (14)$$

To get the intensity, the magnitude of the vector, which is given by multiplication of the complex conjugated form, is squared:

$$I = I_0(E_x E_x^* + E_y E_y^*) = I_0 \left(\frac{\sqrt{2}}{4}\right)^2 [(1 + e^{i\varphi})(1 + e^{-i\varphi}) + (1 + e^{i\varphi})(1 + e^{-i\varphi})]$$
$$= I_0 \frac{1}{8}[4 + 2e^{i\varphi} + 2e^{-i\varphi}] = I_0 \frac{1}{8}[4 + 2(\cos(\varphi) + i\sin(\varphi)) + 2(\cos(\varphi) - i\sin(\varphi))]$$
$$= I_0 \frac{1}{2}[1 + \cos(\varphi)] \quad (15)$$

The transmission through all polarizing elements can be written as:

$$T(\varphi) = \frac{I}{I_0} = \frac{1}{2}[1 + \cos(\varphi)] \quad (16)$$

with only a dependence on the phase retardation of the variable retarder $\varphi = 2\pi R/\lambda$.

By adjusting the term $R = d\rho(\lambda)$, where $d$ represents the thickness of the retarder and $\rho(\lambda)$ is the birefringence of the retarder material, the cosine function can be spectrally tuned and shifted. Therefore, for a given wavelength range, the cosine-shaped transmission can be easily switched to a sine-shaped transmission. If the term $R = d\rho$ is set to 0 (or near zero), the transmission is independent on the wavelength, this position can be used as normalization for the spectral phasor.

As we want to achieve the best fit to a perfect sine and cosine transmission, we also considered the negated sine and cosine modes. For this, the equation (6) was adjusted:

$$I_{sin}(x,y) = \int_{\lambda_{min}}^{\lambda_{max}} I(x,y,\lambda) \cdot q_{sin} \cdot T_{sin}(\lambda) \, d\lambda \quad (17)$$

$$I_{cos}(x,y) = \int_{\lambda_{min}}^{\lambda_{max}} I(x,y,\lambda) \cdot q_{cos} \cdot T_{cos}(\lambda) \, d\lambda \tag{18}$$

The coefficients $q_{sin}$ and $q_{cos}$ can either be +1 or -1 depending on whether the best fit is a positive or negative sine and cosine function in a specific wavelength range. By applying the same coefficient to the $G$ and $S$ component, we would display the points in the phasor plot in a unit circle with start and end at G = 1, S = 0. The correction is given by the following equations:

$$G(x,y) = q_{cos}\left(2\frac{I_{cos}(x,y)}{I_{total}(x,y)} - 1\right) \tag{19}$$

$$S(x,y) = q_{sin}\left(2\frac{I_{sin}(x,y)}{I_{total}(x,y)} - 1\right) \tag{20}$$

Another minor correction was used, due to reason that the used LCVR was not capable of reaching R = 0. Instead, a residual retardance of R =107 nm (at maximum voltage of 25 V, see Fig. S2) resulted in an approx. 6 % lower overall transmission. Since this error was almost constant over the used spectral range (900 to 1600 nm), we applied a correction factor of 1.06 to the $I_{total}(x,y)$ image.

## 3. Spectral unmixing of NIR fluorophores

If a spectrum consists of a mixture of multiple known spectra, this can be used to unmix the concentration of these spectral components. The detected spectrum in each pixel (x,y) is a linear combination of the single spectra of n components:

$$I_n(x,y) = \sum_{k=m}^{n} a_m \, I_m \tag{21}$$

The value $a_m$ represents the fraction of the spectral component $I_m$ inside a pixel. This linear combination also remains valid after the Fourier transformation in the phasor space[1,2]. If there is a mixture of the components $I_1(\lambda)$ and $I_2(\lambda)$, the individual values for $G$ are given by:

$$G_1 = \frac{\int_{\lambda_{min}}^{\lambda_{max}} I_1(\lambda) \cdot T_{cos}(\lambda) \, d\lambda}{\int_{\lambda_{min}}^{\lambda_{max}} I(\lambda) \, d\lambda} \tag{22}$$

$$G_2 = \frac{\int_{\lambda_{min}}^{\lambda_{max}} I_2(\lambda) \cdot T_{cos}(\lambda) \, d\lambda}{\int_{\lambda_{min}}^{\lambda_{max}} I(\lambda) \, d\lambda} \tag{23}$$

where $I(\lambda)$ is the normalization over the sum of both spectral components. For the linear combination thereof, we get:

$$G_{1,2} = \frac{\int_{\lambda_{min}}^{\lambda_{max}} a_1 I_1(\lambda) \cdot T_{cos}(\lambda) \, d\lambda}{\int_{\lambda_{min}}^{\lambda_{max}} I(\lambda) \, d\lambda} + \frac{\int_{\lambda_{min}}^{\lambda_{max}} a_2 I_2(\lambda) \cdot T_{cos}(\lambda) \, d\lambda}{\int_{\lambda_{min}}^{\lambda_{max}} I(\lambda) \, d\lambda} = a_1 G_1 + a_2 G_2 \tag{24}$$

The components are mixed in fractions of $a_1$ and $a_2$, which together are normalized to $a_1 + a_2 = 1$. This simplifies the equation (24) to:

$$G_{1,2} = a_1 G_1 + (1 - a_1) G_2 \tag{25}$$

If we know the pure $G_1$ and $G_2$ values of our sample, we can measure $G_{1,2}$ and unmix the concentrations of the components:

$$a_1 = \frac{G_{1,2} - G_2}{G_1 - G_2} \tag{26}$$

The phasor point for a mixed spectrum of two components is always lying on the connecting line between the individual phasor points. This calculation is analogous for the S values.

**4. Correction for inhomogeneity in macroscopic images**

For macroscopic images the $G$ and $S$ components were corrected by a correction matrix based on a measurement with a white reflectance standard. $I_{cos,corr}(x,y)$, $I_{sin,corr}(x,y)$ and $I_{total,corr}(x,y)$ represent the normalized intensity of the white reflectance standard for the three images. By division of these matrices the images were homogenized, which corrects the uneven illumination and the spatial inhomogeneity of the LCVR. Based on equation (16) we get:

$$G(x,y) = q_{cos} \left(2 \frac{I_{cos}(x,y)/I_{cos,corr}(x,y)}{I_{total}(x,y)/I_{total,corr}(x,y)} - 1\right) \tag{27}$$

$$S(x,y) = q_{sin} \left(2 \frac{I_{sin}(x,y)/I_{sin,corr}(x,y)}{I_{total}(x,y)/I_{total,corr}(x,y)} - 1\right) \tag{28}$$

To avoid repetitive calculations, we defined a correction matrix for $G$ and $S$:

$$G_{corr} = \frac{I_{total,corr}(x,y)}{I_{cos,corr}(x,y)} \tag{29}$$

$$S_{corr} = \frac{I_{total,corr}(x,y)}{I_{sin,corr}(x,y)} \tag{30}$$

Equations (27) and (28) can therefore be simplified as:

$$G(x,y) = q_{cos} \left(2 \, G_{corr} \frac{I_{cos}(x,y)}{I_{int}(x,y)} - 1\right) \tag{31}$$

$$S(x,y) = q_{sin} \left(2 \, S_{corr} \frac{I_{sin}(x,y)}{I_{int}(x,y)} - 1\right) \tag{32}$$

# Supplementary Figures and Tables

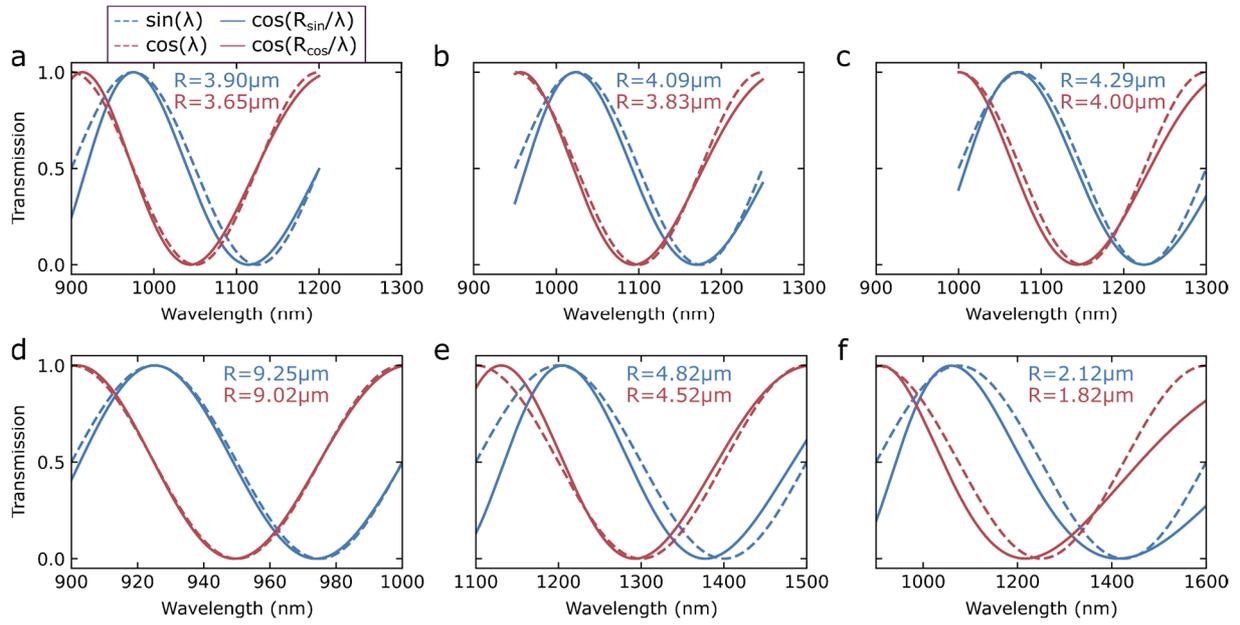

*Fig. S1: Comparison of theoretical (perfect) cos(λ)/sin(λ) versus cos(R/λ) functions.* a) – f) *Transmission functions for different retardances R generated by a variable retarder, which allows the transmission to be precisely matched to a desired spectral range. The value for R was found by fitting the cos(R/λ) function to the perfect cos(λ)/sin(λ) (python package numpy.curve_fit).*

*Tab. S1: Costs to implement HyperNIR imaging (without mechanical parts)*

| High light conditions | | |
|---|---|---|
| Part | Pieces | Costs (€) |
| LPIREA100-C, Thorlabs | 2 | 185.73 |
| LCC1115-B, Thorlabs | 1 | 826.07 |
| KLC101, Thorlabs | 1 | 798.30 |
| | | 1,995.83 |
| **Low light conditions (more efficient transmission components desirable)** | | |
| Part | Pieces | Costs (€) |
| LPNIR100, Thorlabs | 2 | 935.12 |
| LCC1115-B, Thorlabs | 1 | 826.07 |
| KLC101, Thorlabs | 1 | 798.30 |
| | | 3,494.61 |

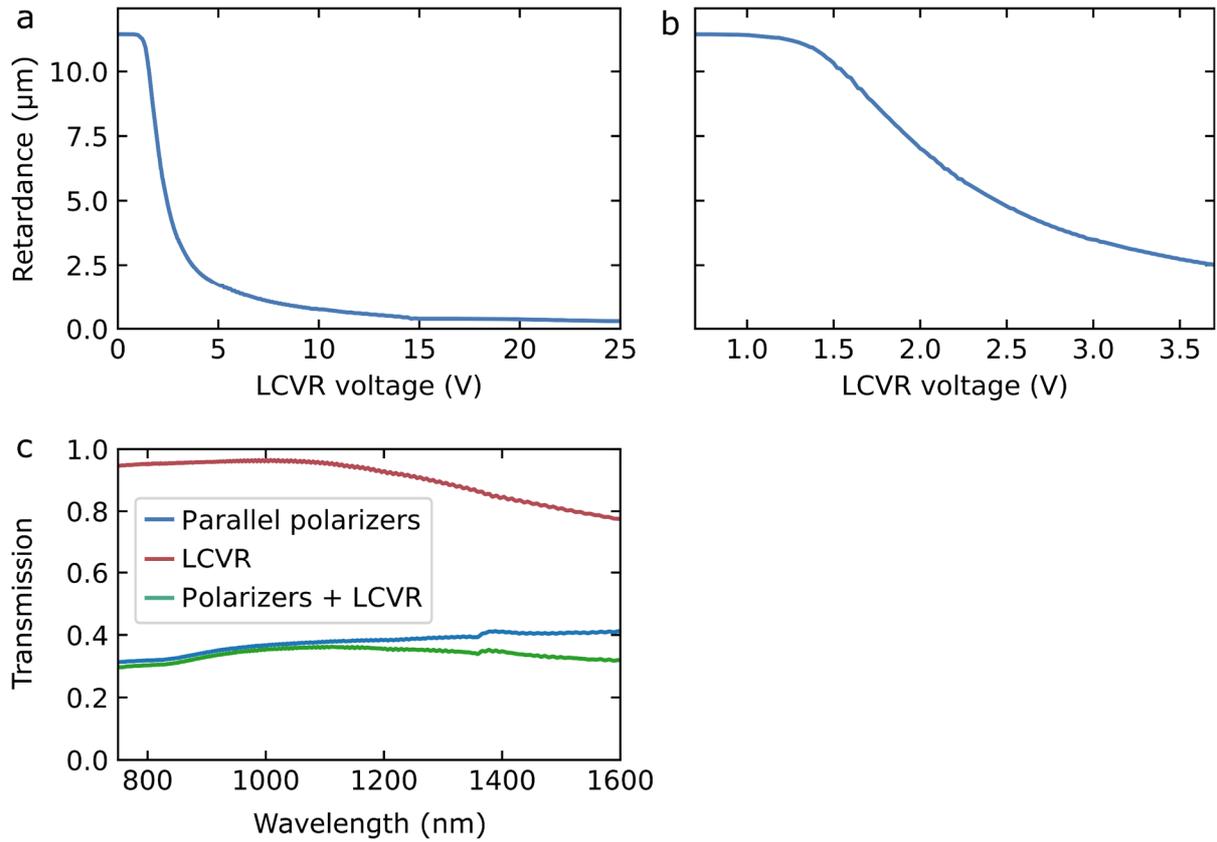

*Fig. S2: Voltage-adjustable retardance.* a) Working curve of the liquid crystal variable retarder (LCVR). By adjusting the LCVR input voltage the retardance can be tuned. b) The biggest retardance change is in the voltage range between 0.8 V and 3.6 V. c) Spectral transmission of the optical components with random polarized light as input. For the complete HyperNIR setup (Polarizers + LCVR) the minimum retardance was selected (R =107 nm at a voltage of 25 V).

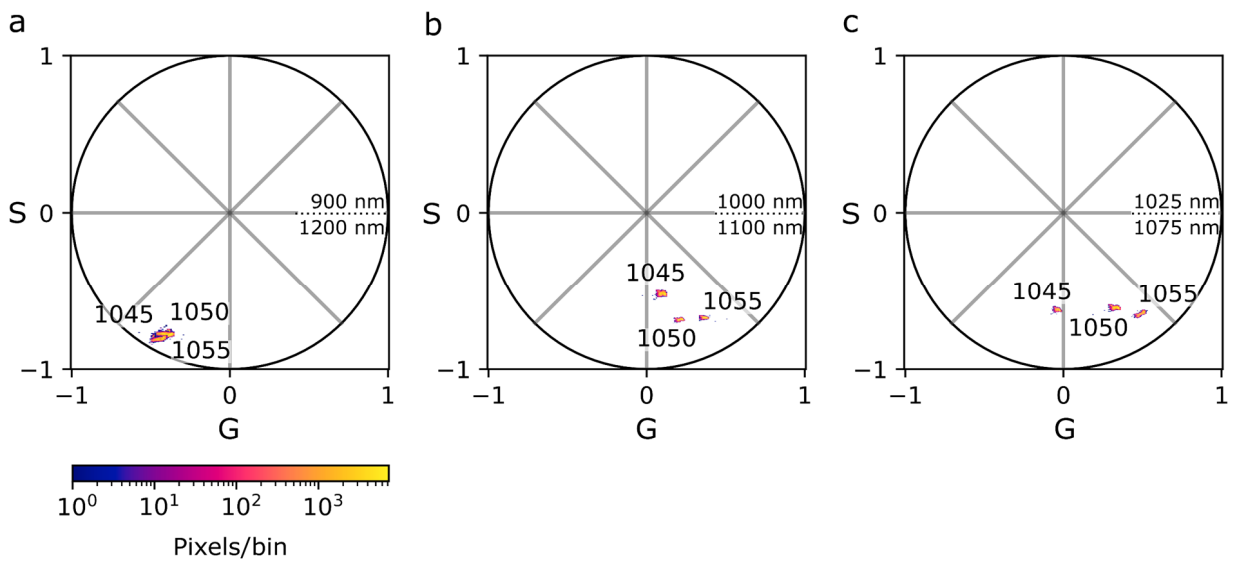

*Fig. S3: Spectral phasors of NIR light signals for different spectral ranges.* Differentiation of spectra with peaks at 1045 nm, 1050 nm and 1055 nm (monochromator with xenon-arc lamp, FWHM = 15 ± 2 nm) for different adjusted spectral ranges / spectral resolution. Ranges: a) 900 – 1200 nm, b) 1000 – 1100 nm, c) 1025 – 1075 nm.

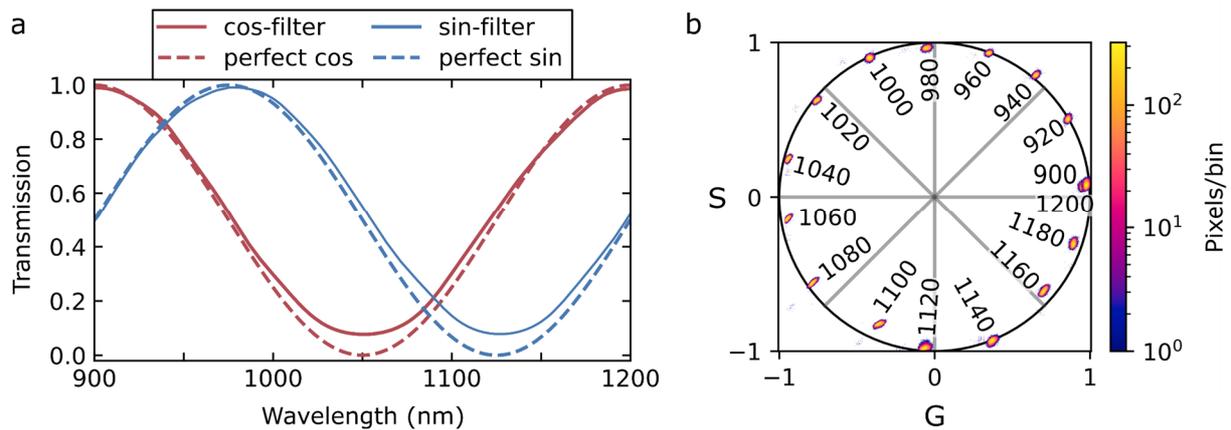

**Fig. S4: NIR phasors with hardware-based sine and cosine spectral filters.** *a) Custom-made interference filters (OptoSigma, costs: 7,754 € for a filter set) designed for the spectral range 900 – 1200 nm to create sine and cosine functions. b) Reference measurement with a monochromatic light source (monochromator with xenon-arc lamp, FWHM = 15 ± 2 nm).*

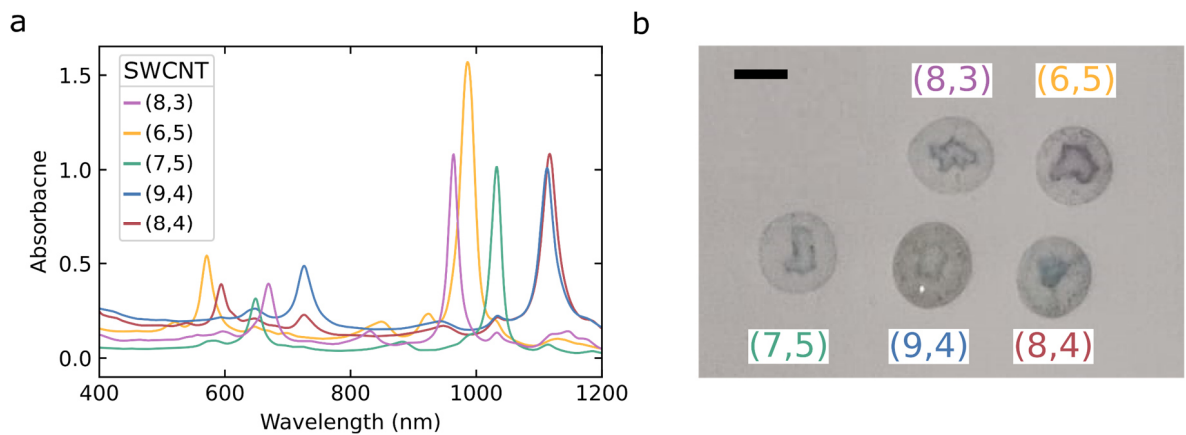

**Fig. S5: NIR fluorescent SWCNT samples.** *a) Absorptions spectra of SWCNTs of different chirality. b) SWCNTs dropcasted on a PVDF membrane imaged with a smartphone camera. Scale bar = 1 mm.*

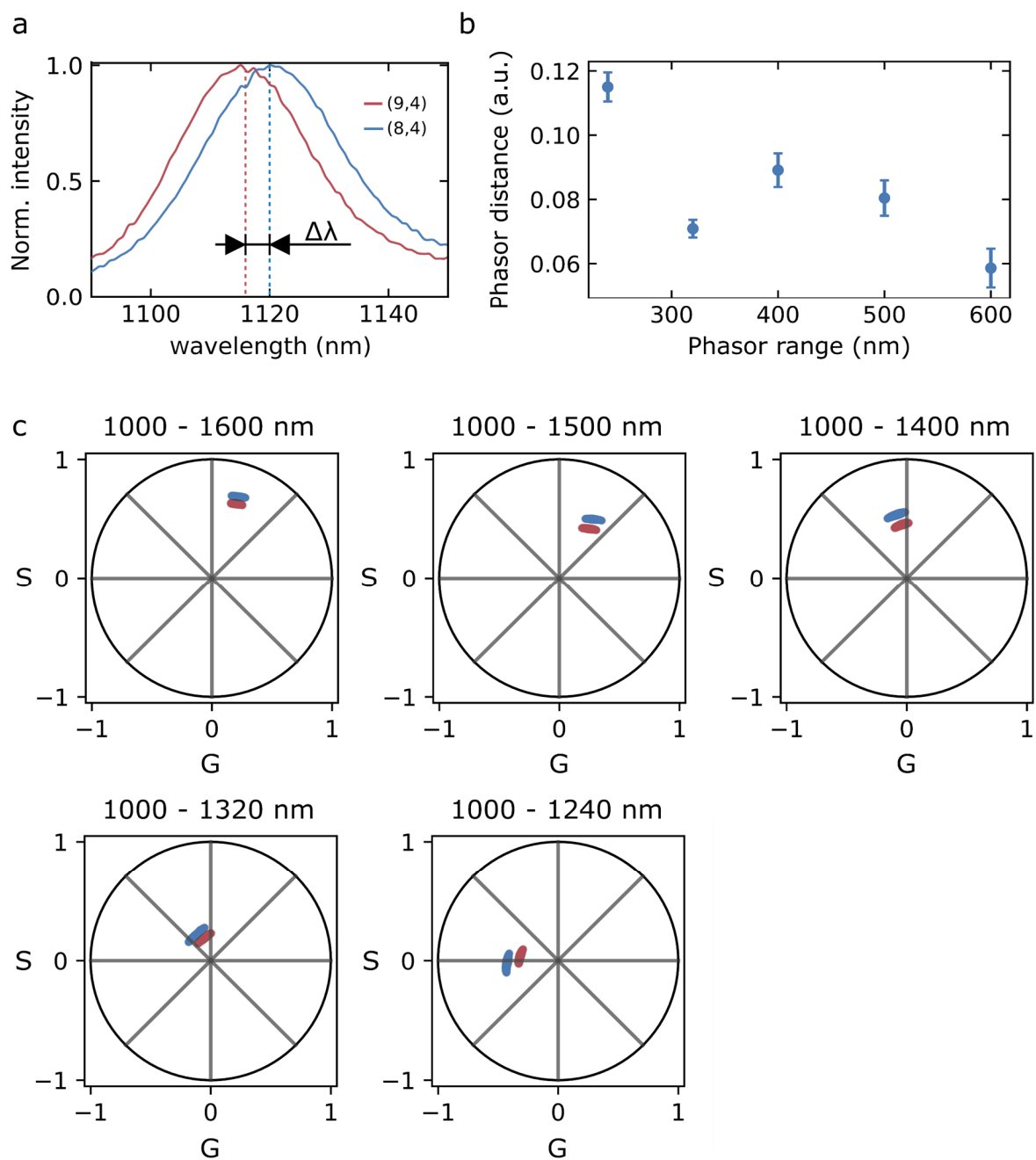

*Fig. S6: Spectral discrimination of similar NIR fluorophores and tuning of spectral resolution.* Differentiation of the (9,4)- and (8,4)-SWCNTs in a phasor plot. a) Emission spectra of the (9,4)- and (8,4)-SWCNTs. b) The calculated mean phasor distance (Euclidian distance) between the phasors of the (9,4)- and (8,4)-SWCNTs for different spectral ranges. Data = Mean ± SD, n = 3. c) Phasor plots of (9,4)- and (8,4)-SWCNTs for these different spectral ranges.

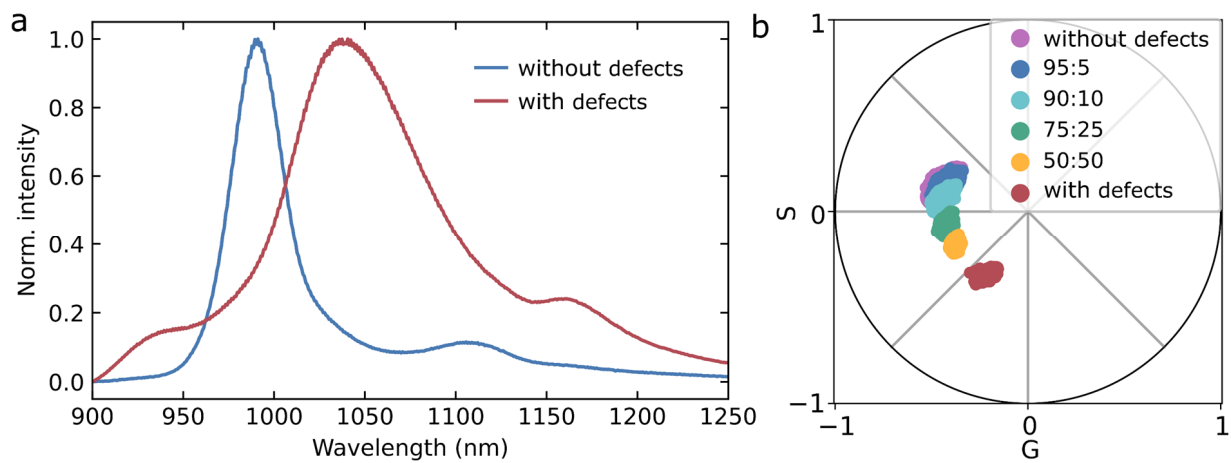

*Fig. S7: Spectral unmixing of SWCNTs with different emission wavelengths based on guanine quantum defects.*
*a) Normalized emission spectra of DNA modified SWCNTs with and without guanine quantum defects. Note that the emission peak position is shifted and the spectrum is broadened when increasing the ratio of SWCNTs with guanine quantum defects. b) Phasor plot for different mixtures of SWCNTs with and without defects in ratios with increasing amounts of SWCNTs without defects. The phasor position can be used to determine the concentration by unmixing.*

*Table S2: Spectral unmixing of mixtures with SWCNTs with and without guanine quantum defects. Data = Mean ± SD, n = 1500 (pixels).*

| Mixture | | Experimentally determined | |
|---|---|---|---|
| Without defects | With defects | Without defects | With defects |
| 50 % | 50 % | 50 ± 3 % | 50 ± 3 % |
| 75 % | 25 % | 71 ± 3 % | 29 ± 3 % |
| 90 % | 10 % | 90 ± 2 % | 10 ± 2 % |
| 95 % | 5 % | 96 ± 2 % | 4 ± 2 % |

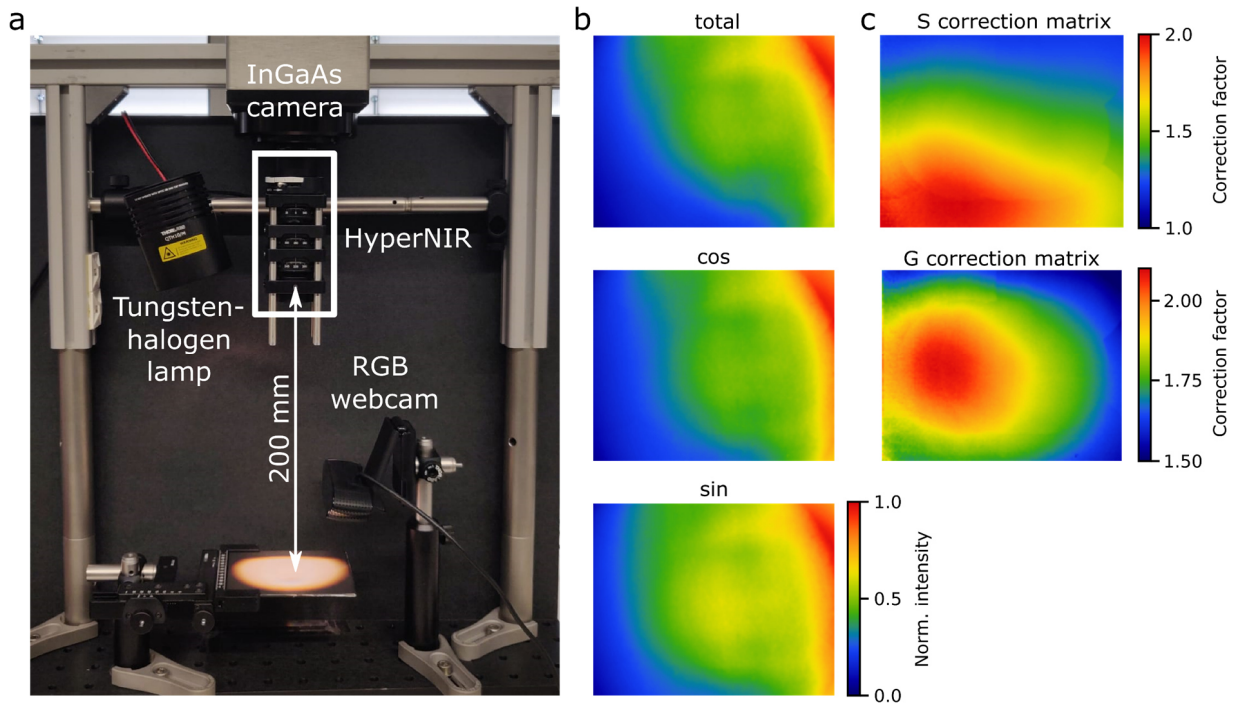

***Fig. S8: Macroscopic reflectance imaging.*** *a) Optical setup. b) Intensity images of a white reflectance standard for the phasor range 900 to 1600 nm. c) Calibration by correction matrices for the G and S component, which are computed by $S_{corr}$ and $G_{corr}$ (see section 4, Eq. (29) & (30)).*

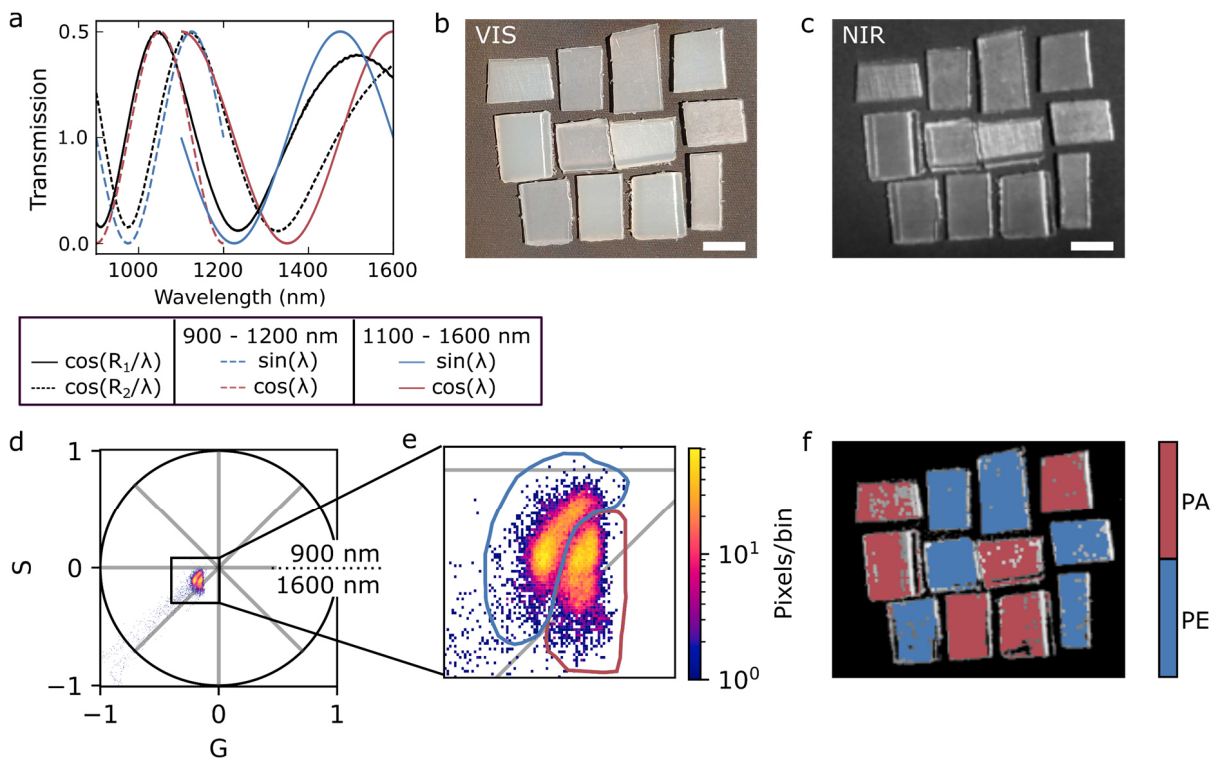

***Fig. S9: Multi-order phasor for plastic differentiation.*** *a) Transmission curve for two different retardances (black). Due to the multi-order cosine wave two different perfect sine and cosine functions are represented (ranges: 900 to 1200 nm and 1100 to 1600 nm). b) Visible image of pieces of PA and PE. c) NIR intensity image. d) Corresponding HyperNIR phasor plot for the range 900 to 1600 nm. e) Magnified plot of the phasor points in (d). f) HyperNIR image (dual-color coding based on the pixel position in the phasor plot according to the blue and red circles in (e). Scale bar = 0.5 mm.*

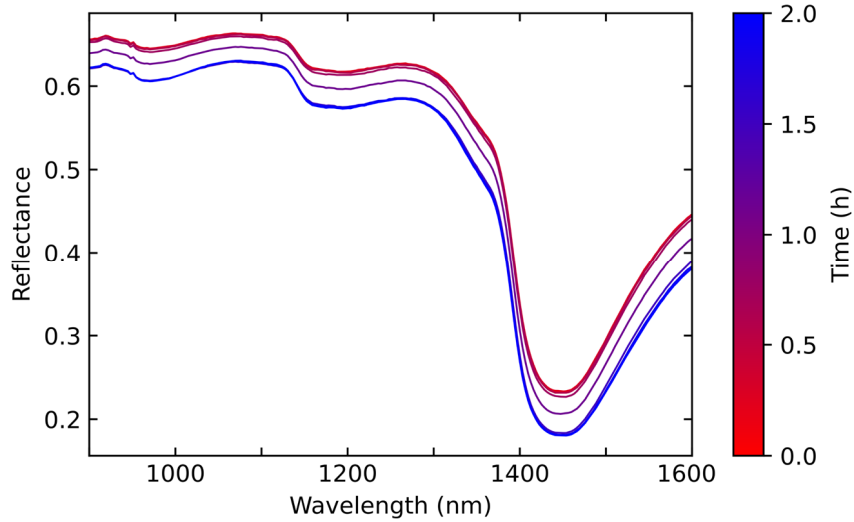

*Fig. S10: NIR spectral variation during water uptake.* The spectral reflectance of a region of interest (focus at the middle of the plant) of a leaf from the pepper plant was measured with an InGaAs spectrometer (AvaSpec-NIR256-1.7-HSC-EVO, Avantes) over 2 h. During the measurement the plant was watered (see Methods section).

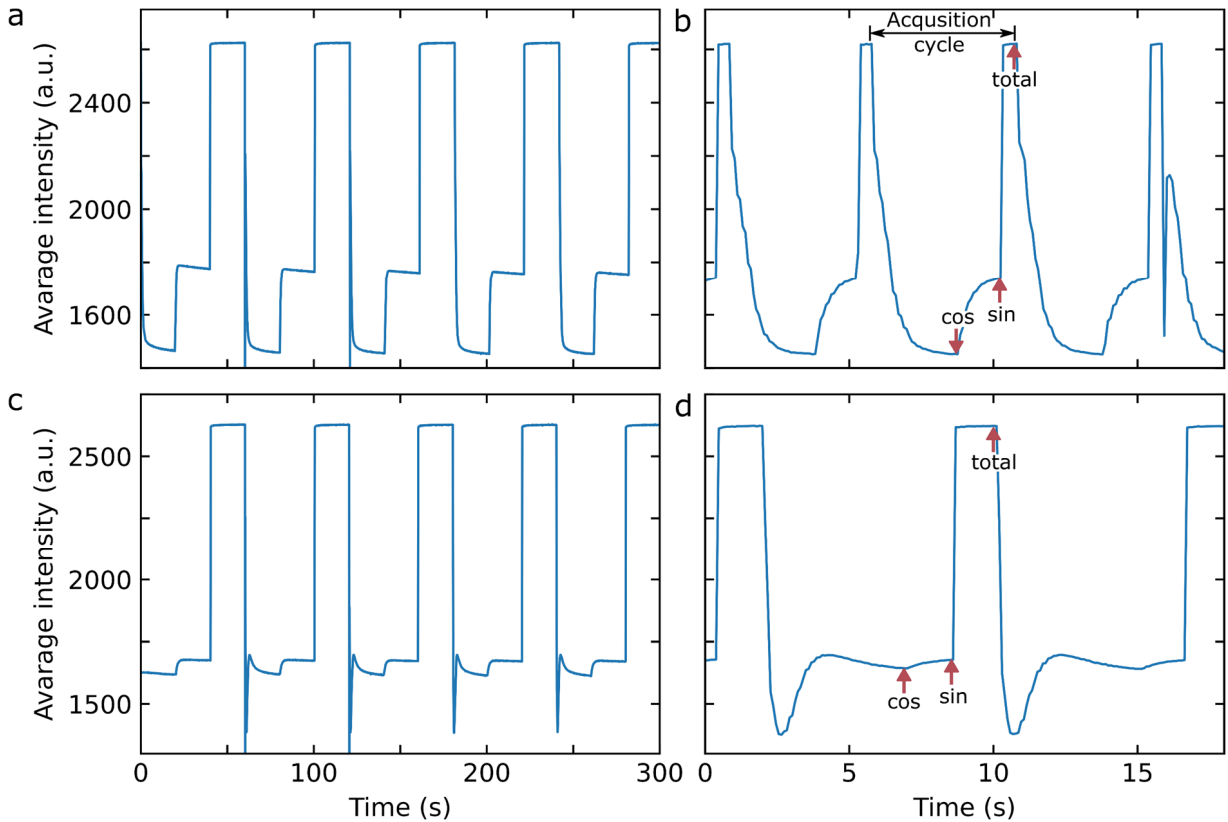

*Fig. S11: Switching time of the LCVR limits video rate.* a) The average in the image during switching of the LCVR driving voltage in the cycle 1.94 V, 2.19 V & 25 V. The time after the next voltage was applied was 10 s. b) Based on the plot in (a) the minimal time until the intensity was steady, was measured and applied as a cycle time for a HyperNIR video. For these voltages the cycle is: set 1.94 V → pause 1.5 s → set 2.19 V → pause 0.5 s → set 25 V → pause 3 s → set 1.94 V. c) The average intensity for a different voltage cycle: 1.35 V, 1.445 V & 25 V. d) Based on (c) the minimal times for the cycle were: set 1.35 V → pause 1.5 s → set 1.445 V → pause 1.5 s → set 25 V → pause 5 s → set 1.35 V. The time points at which the three images (cos, sin, total) were taken are marked.

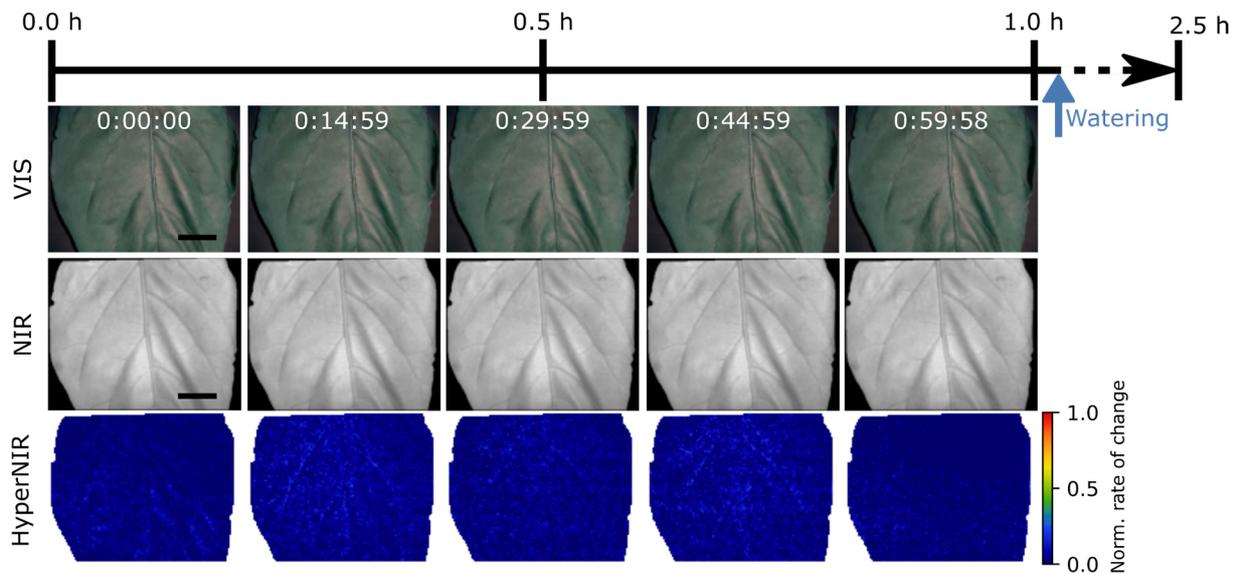

*Fig. S12: Dry leaf measured over a duration of 1 h.* Measurement of the time before the measurement for water uptake was started. The Fig. 5b (main manuscript) shows the following 1.5 h of this measurement (see also Supplementary Video 1). Note that the normalized rate of change is the relative change in the Euclidian distance in phasor space for each pixel. Normalization of the rate of change was calculated similar to Fig. 5b. Scale bar = 1 mm.

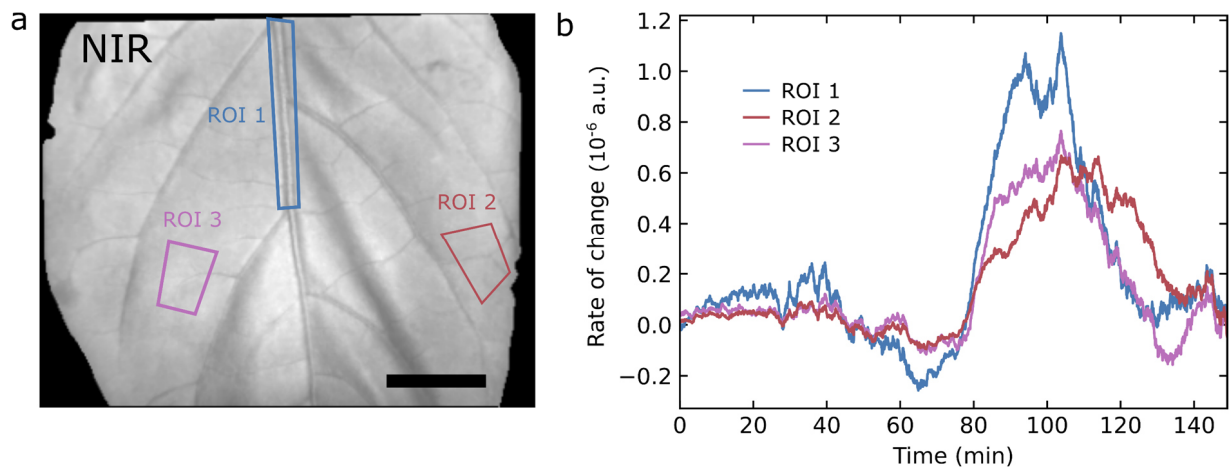

*Fig. S13: Phasor changes in response to water uptake for different regions of interests in a leaf.* a) Leaf with three regions of interests: ROI 1 is the main capillary of this leaf, ROI 2 and ROI 3 are different tissue parts, which represent the area around a small capillary. b) The mean rate of change in the different ROIs was calculated for every time frame of the video (Supplementary Video 1). Note that the rate of change is the relative change in the Euclidian distance in phasor space for each pixel. Scale bar = 1 mm.

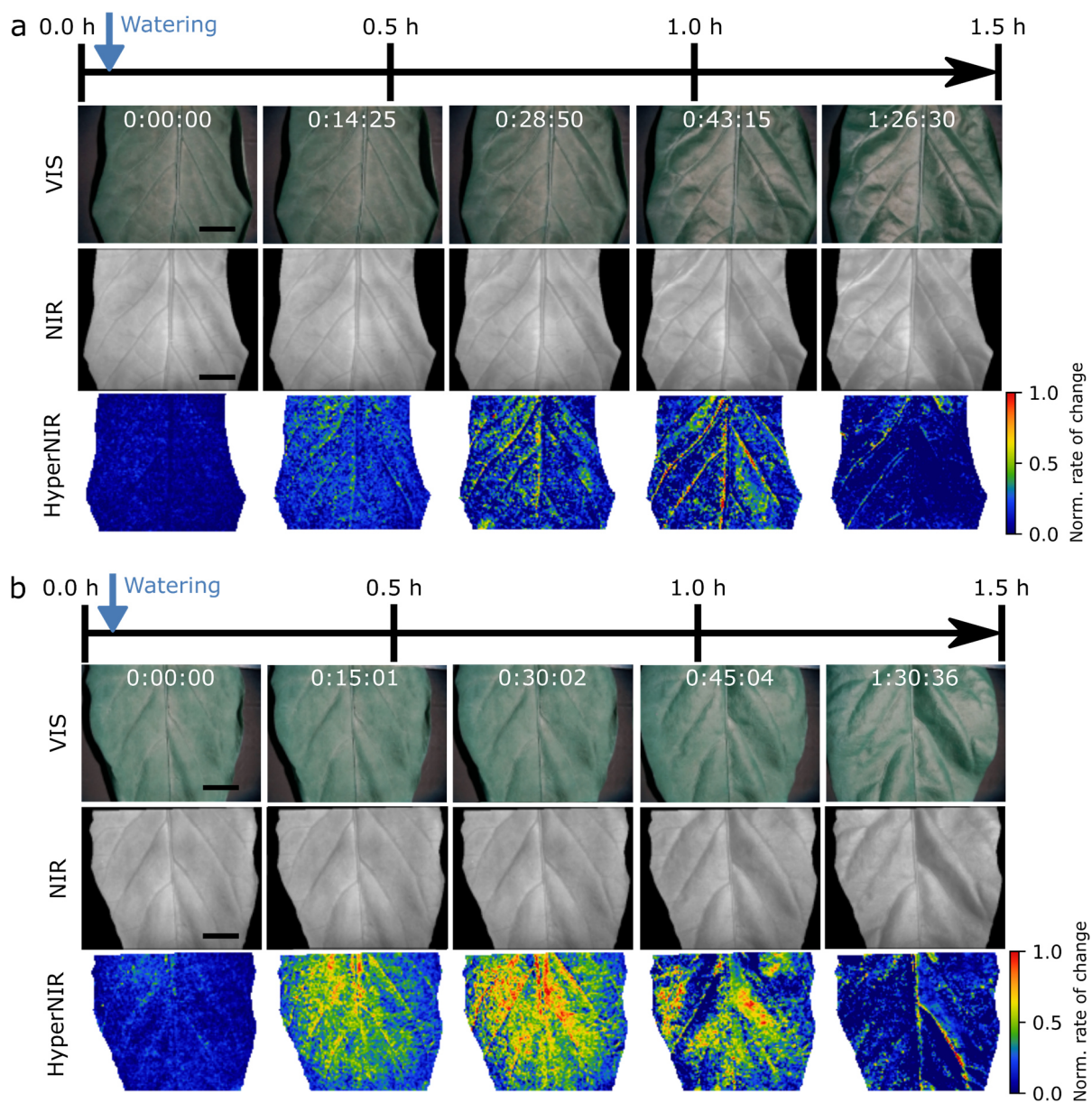

*Fig. S14: **HyperNIR imaging of water uptake in leaves.** After approx. 1 min the plant was watered. In the HyperNIR images the normalized rate of change is visualized, which represents the movement of the phasor relative to the phasor position at the beginning for each pixel. a) and b) are different leaves. Note that the normalized rate of change is the relative change in the Euclidian distance in phasor space for each pixel. Scale bar = 1 mm.*